\documentclass[a4paper,11pt,pdftex]{article}
\usepackage{jheppub} 
\usepackage[utf8]{inputenc}
\usepackage{amsfonts}
\usepackage{amsbsy}
\usepackage{url}
\usepackage{enumerate}
\usepackage[justification=raggedright]{caption}
\usepackage{hhline}

\usepackage{graphicx,color}
\usepackage{amssymb,amsmath}
\usepackage{slashed}
\usepackage{hyperref}
\usepackage{braket}
\usepackage{simplewick}
\usepackage{caption}

\usepackage{ascmac}
\usepackage{latexsym}
\usepackage{pifont}          
\usepackage{bm}
\usepackage{comment}

\newcommand{\n}{\nonumber}

\newcommand{\mr}[1]{\mathrm{#1}}

\newcommand{\h}[1]{\hspace{#1}}
\newcommand{\f}[2]{\frac{#1}{#2}}

\newcommand{\ol}[1]{\tilde{#1}}

\makeatletter
\@addtoreset{equation}{section}
\makeatother

\begin{document}
\preprint{KUNS-2838}

\title{Electroweak axion string and superconductivity}

\author{Yoshihiko~Abe,}
\author{Yu~Hamada,}
\author{and Koichi~Yoshioka}

\affiliation{Department of Physics, Kyoto University, Kitashirakawa, Kyoto 606-8502, Japan}

\emailAdd{y.abe(at)gauge.scphys.kyoto-u.ac.jp}
\emailAdd{yu.hamada(at)gauge.scphys.kyoto-u.ac.jp}
\emailAdd{yoshioka(at)gauge.scphys.kyoto-u.ac.jp}

\abstract{
We study the axion strings
with the electroweak gauge flux in the DFSZ axion model
and show that these strings, called the electroweak axion strings, can exhibit superconductivity
without fermionic zeromodes.
We construct three types of electroweak axion string solutions.
Among them, the string with $W$-flux can be lightest in some parameter space,
which leads to a stable superconducting cosmic string.
We also show that a large electric current can flow along the string due to
the Peccei-Quinn scale much higher than the electroweak scale.
This large current induces a net attractive force between the axion strings
with the same topological charge, 
which opens a novel possibility that 
the axion strings form Y-junctions in the early universe.
}

\maketitle

\section{Introduction}

The strong CP problem is one of the unresolved mysteries in the Standard Model (SM) of particle physics.
The problem can be naturally solved by the Peccei-Quinn mechanism,
in which a global symmetry denoted by $U(1)_\mathrm{PQ}$ is assumed to be spontaneously broken 
and provides a (pseudo) Nambu-Goldstone particle, the axion \cite{Peccei:1977ur,Peccei:1977hh,Weinberg:1977ma,Wilczek:1977pj}.
The axion is a promising candidate for a viable cold dark matter \cite{Preskill:1982cy,Abbott:1982af,Dine:1982ah}.

Among various models bringing the axion (for recent reviews, see, e.g., Refs.~\cite{Sikivie:2006ni,Marsh:2015xka,Ringwald:2012hr,Wantz:2009it,Kim:2008hd}),
 the DFSZ model \cite{Zhitnitsky:1980tq,Dine:1981rt} has been studied extensively,
as well as the KSVZ model \cite{Kim:1979if,Shifman:1979if}.
In the DFSZ model, the scalar sector of the SM is extended to have two Higgs doublets 
and one SM-singlet complex scalar. 
The scalar fields and the SM fermions are assumed to be charged under the $U(1)_\mathrm{PQ}$ symmetry,
which is spontaneously broken by a vacuum expectation value (VEV) of the complex scalar.
The axion is a linear combination of imaginary components of the doublets and the complex scalar.

The $U(1)_\mathrm{PQ}$ symmetry in the DFSZ model is anomalous due to one-loop contributions from the SM fermions
and is broken down to a discrete subgroup $\mathbb{Z}_3$ (or $\mathbb{Z}_6$),
which produces a domain wall at the QCD phase transition.
The energy density of the domain walls dominate soon that of the universe,
leading to the domain wall problem.
One possible scenario to solve the problem is to assume that the $U(1)_\mathrm{PQ}$ symmetry is broken during or before the cosmological inflation.
There exists, however, a stringent constraint on isocurvature perturbation produced by the axion during the inflation \cite{Akrami:2018odb}.
Another simple way is introducing a tiny term breaking the symmetry explicitly, called the bias \cite{Gelmini:1988sf,Larsson:1996sp},
which resolves the vacuum degeneracy \cite{Sikivie:1982qv,Chang:1998tb,Vilenkin:1981zs}.
For other scenarios and their studies, see, e.g., Refs.~\cite{Peccei:1986pn,Krauss:1986wx,Lazarides:1982tw,Chatterjee:2019rch,Kawasaki:2015lpf,Sato:2018nqy}.

As well as most axion models enjoying the $U(1)_\mathrm{PQ}$ symmetry, 
the DFSZ model predicts the axion string \cite{Davis:1986xc}, which is a global cosmic string.
The axion string is created by the Kibble-Zurek mechanism \cite{Kibble:1980mv,Zurek:1985qw} (see also Ref.~\cite{Murayama:2009nj})
when the $U(1)_{\mathrm{PQ}}$ symmetry is spontaneously broken.
For the scenario that the domain wall problem is avoided by the inflation, 
the axion strings are diluted away and seem to play no role in cosmology.
But for the other scenarios, they become interesting ingredients in the universe 
and have been studied in various contexts (see Ref.~\cite{Vilenkin:2000jqa}). 
We assume the latter scenarios in this paper.
After the creation, the strings form a network whose energy density has a scaling property.
To understand the evolution of the network, it is important to study the interaction between the axion strings.
The interaction is thought to be dominated by exchange of the (massless) axion as a long-range force.

On the other hand, cosmic strings sometimes can be superconducting strings \cite{Witten:1984eb}
when the electromagnetic gauge symmetry is spontaneously broken inside the strings.
It is known that the axion strings necessarily become superconducting states \cite{Lazarides:1984zq,Iwazaki:1997bk,Ganoulis:1989hz,Lazarides:1987rq} because they must have fermionic zero modes traveling on the string \cite{Jackiw:1981ee,Callan:1984sa}.
The maximum amount of the supercurrent is determined by the (bulk) mass of the fermions.
In the DFSZ model, however, the axion string cannot carry significant amount of the current because the model has no heavy fermion.
Thus, the superconductivity seems to play no crucial role for cosmological properties.

In this paper, we show that 
the axion string in the DFSZ model becomes the electroweak string
after the breaking of the electroweak symmetry.
The electroweak string is a string containing flux tubes of the $SU(2)_W \times U(1)_Y$ gauge fields like the Abrikosov-Nielsen-Olesen vortex \cite{Abrikosov:1956sx,Nielsen:1973cs}
and has been studied in the SM \cite{Nambu:1977ag,Vachaspati:1992jk,Vachaspati:1992fi,James:1992zp,James:1992wb,Vachaspati:1994ng,Barriola:1993fy,Barriola:1994ez,Eto:2012kb}
(see Ref.~\cite{Achucarro:1999it} for review),
and in two Higgs doublet models (2HDM) \cite{Perivolaropoulos:1993gg,La:1993je,Dvali:1993sg,Dvali:1994qf,Bimonte:1994qh,Bachas:1998bf,Ivanov:2007de,Battye:2011jj,Eto:2019hhf,Eto:2020hjb} (for recent comprehensive studies, see Refs.~\cite{Eto:2018hhg,Eto:2018tnk}).
An essence is that the two Higgs doublets in the DFSZ model also acquire the VEVs after the electroweak phase transition
and they must have winding in the $SU(2)_W \times U(1)_Y$ gauge orbits for the single-valuedness, as well as the winding for $U(1)_\mathrm{PQ}$.
We call such strings \textit{the electroweak axion strings}.
In particular, we show that there are at least three types of the electroweak axion string in the DFSZ model.
Interestingly, some of them have very similar properties to those of the electroweak string in 2HDM.

Furthermore, we show that one of the electroweak axion strings (dubbed the type-C string) can be a superconducting string \textit{without} fermionic fields.
This is because the charged fields, the charged Higgs and $W$ bosons, acquire non-zero values inside the string and the $U(1)_\mathrm{EM}$ symmetry is spontaneously broken there.
This is a similar situation to superconductivity of non-Abelian vortices \cite{Alford:1990mk,Alford:1990ur} and of the $U(1)\times \tilde{U}(1)$ model considered by Witten \cite{Witten:1984eb}.
Remarkably, due to the coupling between the Higgs doublets and the complex scalar, 
the amount of the supercurrent can be of order of the $U(1)_\mathrm{PQ}$ breaking scale resulting in large magnetic energy even in the DFSZ model.
As a consequence, the strings feel a large magnetic interaction, which can overcome the one from the axion exchange.
Therefore, superconductivity could drastically change the cosmological scenario of the axion strings 
after the electroweak phase transition in the DFSZ model.

The rest of this paper is organized as follows.
In Sec.~\ref{005936_1Oct20},
the DFSZ axion model is reviewed and our notation is introduced.
For later use, we present a definition of the $U(1)_\mathrm{EM}$ in general soliton backgrounds.
In Sec.~\ref{sec:electroweakaxionstrings},
after a brief review of the conventional axion string,
we discuss the electroweak axion strings.
There are at least three types of the electroweak axion strings (type-A, B and C).
We compare the tensions of the strings.
In Sec.~\ref{sec:superconducting},
we show that the type-C string can be superconducting.
A linearized equation of motion for massless zero modes traveling on the string is presented.
In addition, we estimate the maximum amount of the supercurrent flowing on the string
to be of order of the $U(1)_\mathrm{PQ}$ breaking scale.
Sec.~\ref{sec:conclusion} is devoted to the conclusion.
In Appendix.~\ref{041710_25Sep20}, we present the derivation of the linearized equation used in Sec.~\ref{sec:superconducting}.

\section{The model}
\label{005936_1Oct20}

\subsection{DFSZ axion model}

\begin{table}[t]
\centering
\begin{tabular}{|c||c|c|c|}\hline
 &$H_1$&$H_2$&$S$\\
 \hhline{|=#=|=|=|}
 $SU(2)_W$&$\bm{2}$&$\bm{2}$&$\bm{1}$ \\ \hline
 $U(1)_Y$ &$1$&$1$&$0$ \\ \hline
 $U(1)_{\mathrm{\mathrm{PQ}}}$&$X_1$&$X_2$&$X_s$\\ \hline
\end{tabular}
\caption{
The scalar field contents and their quantum charges.
}
\label{tab:contents}
\end{table}

The particle contents and the charge assignments under the SM gauge group and the $U(1)_{\mathrm{\mathrm{PQ}}}$
are shown in Tab.~\ref{tab:contents}.
We introduce a SM-singlet complex scalar $S$ and two $SU(2)_W$ doublets, $H_1$ and $H_2$, both with the hypercharge $Y=1$.
The Lagrangian which describes the electroweak and scalar sectors is written as
\begin{align}
\h{-1em} {\mathcal L} = - \frac{1}{4}\left(Y_{\mu\nu}\right)^2 - \frac{1}{4}\left(W_{\mu\nu}^a\right)^2  
+ \sum_{i=1,2} \left|D_\mu H_i \right|^2 + \left|\partial_\mu S \right|^2 - V(H_1, H_2, S).
\label{eq:L}
\end{align}
Here, $Y_{\mu\nu}$ and $W^a_{\mu\nu}$ describe field strength tensors of the hypercharge
and weak gauge interactions, respectively, with $\mu$ ($\nu$) and $a$ being Lorentz and weak iso-spin indices, respectively. 
$D_\mu$ represents the covariant derivative acting on the Higgs fields, and the index $i$ runs $i=1,2$.
The scalar potential $V(H_1, H_2,S)$ being invariant under the charge assignments of Tab.~\ref{tab:contents}
is
\begin{align}
V(H_1,H_2,S) & = V_H + V_S + V_\mathrm{mix},
\end{align}
where each part is given by
\begin{align}
 V_H  =&  m_{11}^2H_1^\dagger H_1 + m_{22}^2 H_2^\dagger H_2 
 + \frac{\beta_1}{2}\left(H_1^\dagger H_1\right)^2 + \frac{\beta_2}{2}\left(H_2^\dagger H_2\right)^2
 \n\label{213918_10Sep20}
 \\
 &+ \beta_3\left(H_1^\dagger H_1\right)\left(H_2^\dagger H_2\right) 
 + \beta_4 \left(H_1^\dagger H_2\right)\left(H_2^\dagger H_1\right),
 \\
 V_S =& - m_S^2 |S|^2 + \lambda_S |S|^4,
 \\
 V_\mathrm{mix}=& \left( \kappa S^2 H_1^\dagger H_2 + \mathrm{h.c.} \right) 
 + \kappa_{1S} |S|^2 |H_1|^2 +  \kappa_{2S} |S|^2 |H_2|^2,\label{171916_6Oct20}
\end{align}
with $m_S^2 > 0$ which admits $S$ to acquire a non-zero VEV: 
$\braket{S} = v_s$.
Without loss of generality,
we can suppose that the Higgs fields develop VEVs as
$ \Braket{H_1} = \left( 0,  v_1\right)^{\mathrm{T}},~\Braket{H_2} =  \left(0 , v_2 \right)^{\mathrm{T}} $ with $v_1,v_2 \in \mathbb{R}$.\footnote{
Note that we drop ``$1/\sqrt{2}$'' in our notation for the VEVs.}
Then the electroweak scale, $v_\mr{EW}$ ($\simeq $ 246 GeV), can be expressed by these VEVs as 
$v_{\rm EW}^2 = 2 (v_1^2 + v_2^2)$.
We also define $\tan \beta \equiv v_2/ v_1$.
In order for $V_\mathrm{mix}$ to be invariant under the $U(1)_\mathrm{PQ}$ symmetry, 
the $U(1)_{\mathrm{\mathrm{PQ}}}$ charges in Tab.~\ref{tab:contents} should satisfy the relation $2 X_s - X_1 + X_2 =0$.
If the first term in Eq.~\eqref{171916_6Oct20} has a structure like $S H_1 ^\dagger H_2$ instead of $S^2 H_1 ^\dagger H_2$,
the assignment of the $U(1)_\mathrm{PQ}$ charges should change,
but qualitative properties of the axion strings we discuss below are almost same.
In particular, the string becomes superconducting also in such a case.

The Yukawa interaction terms are given by
\begin{align}
 \mathcal{L}_{\mathrm{Yukawa}} =&
 - y_U \overline{Q} \bigl( i \sigma_2 H_1^* \bigr) u_R
 - y_D \overline{Q} H_2 d_R
 - y_{e} \overline{L} H_2 e_R
 + \mathrm{h.c.},
\end{align}
and the SM fermions carry the $U(1)_{\mathrm{\mathrm{PQ}}}$ charge
so that this Lagrangian is invariant under $U(1)_{\mathrm{\mathrm{PQ}}}$.
The new singlet scalar $S$ couples to the SM fermions via Higgs sector.
In the following parts of this paper,
we leave aside the Yukawa terms.

For later use, we rewrite the Higgs fields in a two-by-two matrix form\cite{Grzadkowski:2010dj}, $H$,
defined by 
\begin{equation}
 H = \left( i\sigma_2 H_1^*,\ H_2\right).
\end{equation}
The matrix field $H$ transforms under the electroweak $SU(2)_W \times U(1)_Y$ symmetry as
\begin{equation}
H \to \exp\left[\f{i}{2}\theta_a(x) \sigma_a\right] H ~\exp\left[-\f{i}{2} \theta_Y(x) \sigma_3\right],
\end{equation}
where the group element acting from the left belongs to $SU(2)_W$ and the other element acting from the right belongs to $U(1)_Y$.
Therefore the covariant derivative on $H$ can be expressed as
\begin{equation}
D_\mu H =\partial_\mu H - i \frac{g}{2} \sigma_a W_\mu^a H + i \frac{g'}{2}H\sigma_3 Y_\mu.
\end{equation}
The VEV of $H$ is expressed by a diagonal matrix $\langle H \rangle = \mr{diag} (v_1,v_2)$,
and the Higgs potential $V_H$ can be written by using $H$ as follows:
\begin{align}
V_H
 =& - m_{1}^2~ \mr{Tr}|H|^2 - m_{2}^2~ \mr{Tr}\left(|H|^2 \sigma_3\right)
 + \alpha_1~\mr{Tr}|H|^4 \n\label{213928_10Sep20} \\
 & +  \alpha_2 ~\left(\mr{Tr}|H|^2 \right)^2
 + \alpha_3~ \mr{ Tr}\left(|H|^2 \sigma_3 |H|^2\sigma_3\right)  
 + \alpha_4~ \mr{Tr}\left(|H|^2 \sigma_3 |H|^2\right),
\end{align}
where $|H|^2 \equiv H ^\dagger H$
and the relations between the parameters in Eq.~\eqref{213918_10Sep20} and in Eq.~\eqref{213928_10Sep20} are given by 
\begin{align}
&  m_{11}^2 = -m_1^2 - m_2^2 , \h{2em}  m_{22}^2 = -m_1^2 + m_2^2 ,\\ 
&  \beta_1  =2(\alpha_1+ \alpha_2 + \alpha_3 +  \alpha_4 ), \h{2em} \beta_2  =2( \alpha_1+ \alpha_2 + \alpha_3 -  \alpha_4 ),
\label{eq:beta1}
\\
&  \beta_3 = 2(\alpha_1+ \alpha_2 - \alpha_3) ,  \h{2em} \beta_4 = 2(\alpha_3 - \alpha_1).
\label{eq:beta2}
\end{align}
The mixing term $V_\mathrm{mix}$ is also rewritten by $H$ as
\begin{align}
  V_\mathrm{mix} =& \left( \kappa S^2 \mathrm{det} H + \mathrm{h.c.} \right) + \frac{1}{2}(\kappa_{1S}+\kappa_{2S}) |S|^2 ~\mathrm{Tr}|H|^2 \n\label{183300_19Sep20} \\
&+ \frac{1}{2}(\kappa_{1S}-\kappa_{2S}) |S|^2 ~\mathrm{Tr}(|H|^2 \sigma_3).
\end{align}

The custodial transformation $SU(2)_C$ in the two Higgs doublet model \cite{Grzadkowski:2010dj,Pomarol:1993mu}
is identified as the global unitary transformation of the Higgs matrix $H$ as
\begin{align}
 H \to U^\dagger H U,
 \quad
 U \in SU(2)_C.
\end{align}
If $m_2^2 =\alpha_3 = \alpha_4 =0$, and $\kappa_{1S} = \kappa_{2S}$,
the scalar potential is invariant under the custodial transformation.
This symmetry makes the two VEVs be equal, $\tan \beta=1$.


\subsection{Mass spectra and PQ transformation}

The scalar fields develop the following VEVs
\begin{align}
 \Braket{H} = \begin{pmatrix}
 v_1 & 0 \\
 0 & v_2
 \end{pmatrix},
 \quad
 \Braket{S} = v_s.
\end{align}
The stationary conditions are solved with the mass parameters 
$m_1^2$, $m_2^2$, $m_S^2$ as
\begin{align}
 m_1^2 &=  (\alpha_1 + 2 \alpha_2 + \alpha_3 + \alpha_4) v_1^2 
 + (\alpha_1 + 2 \alpha_2 + \alpha_3 - \alpha_4) v_2^2
 + \frac{1}{2}\biggl[ \kappa \biggl( \frac{v_1}{v_2} + \frac{v_2}{v_1} \biggr) + \kappa_{1S} + \kappa_{2S} \biggr]
 v_s^2,
 \\
 m_2^2 &= ( \alpha_1 + \alpha_3 + \alpha_4) v_1^2
 - (\alpha_1 + \alpha_3 - \alpha_4) v_2^2
 + \frac{1}{2} \biggl[ \kappa \biggl(\frac{v_2}{v_1} - \frac{v_1}{v_2} \biggr) + \kappa_{1S} - \kappa_{2S} \biggr]
 v_s^2,\label{143600_6Oct20}
 \\
 m_S^2 &= 2 \lambda_S v_s^2 + \kappa_{1S} v_1^2 + \kappa_{2S} v_2^2 + 2 \kappa v_1 v_2.
\end{align}
After the symmetry breaking,
the gauge bosons and scalars become massive due to the above VEVs.
The masses of the weak gauge bosons are given by
\begin{align}
 m_W = \frac{gv_{\rm EW}}{2},
 \quad
 m_Z = \frac{gv_{\rm EW}}{2\cos\theta_W},
\end{align}
with the standard definitions of the weak mixing angle
$\cos\theta_W = g/\sqrt{g^2 + g'{}^2}$, the $Z$ boson
$Z_\mu = W_\mu ^3 \cos\theta_W - Y_\mu \sin\theta_W$,
and the photon
$A_\mu = W_\mu ^3 \sin\theta_W + Y_\mu \cos\theta_W$.

In the scalar sector, we have three scalars, three pseudo scalars, and
two charged scalars. Among these,
one massless pseudo scalar and one massless charged scalar are eaten by the
weak gauge bosons,
and the other massless pseudo scalar becomes the axion,
which obtains a mass from the non-perturbative QCD effect.
There remain five physical scalar bosons after the symmetry breaking.
For example, the lightest real scalar has the mass eigenvalue 
\begin{align}
 m_{h_1}^2 &\simeq 4 (\alpha_1 + \alpha_3) \frac{v_1^4 + v_2^4}{v_1^2 + v_2^2} + 4 \alpha_2 (v_1^2 + v_2^2)
 + 4 \alpha_4 (v_1^2 -v_2^2)
 \nonumber\\
 & \h{2em}- \frac{\bigl( \kappa_{1S} v_1^2 + \kappa_{2S} v_2^2 + 2 \kappa v_1 v_2 \bigr)^2}{\lambda_S (v_1^2 + v_2^2) }
 ,
\end{align}
up to $\mathcal{O}(v_{\mathrm{EW}}^2/ v_s^2)$, and we 
identify it as the SM Higgs boson. 
The mass squared matrix for three pseudo scalars is expressed as
\begin{align}
 \kappa \left( \begin{array}{ccc}
 - \frac{v_2 v_s^2}{v_1} & v_s^2 & 2 v_2 v_s \\
 v_s^2 & - \frac{v_1 v_s^2}{v_2} & - 2 v_1 v_s\\
 2 v_2 v_s & -2 v_1 v_s & -4 v_1 v_2
 \end{array}\right).
\end{align}
This matrix has one massive and two exact zero modes. 
The non-vanishing mass eigenvalue is
\begin{align}
 m_{A_0}^2 = - \kappa \frac{4 v_1^2 v_2^2 + v_1^2 v_s^2 + v_2^2 v_s^2}{v_1 v_2}.
\end{align}
In order to avoid the tachyonic mass,
the portal coupling $\kappa$ should be negative.
One of the
massless eigenvector is $(\cos\beta,\sin\beta,0)$ which corresponds to
the longitudinal mode of the $Z$ boson.
Another zero eigenvector is given by $(X_1v_1,X_2v_2,X_s v_s)$ as long as
$2X_s-X_1+X_2=0$ is satisfied. This flat direction corresponds to the
axion. Imposing these two massless modes are orthogonal, we find
\begin{align}
 X_1 =  2 \sin^2 \beta,
 \quad
 X_2 = - 2 \cos^2 \beta,
 \quad
 X_s =1,
\end{align}
where $X_s$ determines the normalization. 
Then the $U(1)_{\mathrm{PQ}}$ transformation acts on the scalars as
\begin{align}
 H_1 \to  e^{2 i \alpha \sin^2 \beta} H_1, \h{1em}
 H_2 \to e^{-2 i \alpha \cos^2 \beta} H_2, \h{1em}
 S \to e^{i \alpha} S. 
 \label{224951_10Sep20}
\end{align}
The same result is obtained by 
defining the $U(1)_{\mathrm{PQ}}$ current not to couple to the $Z$ boson.
For the matrix field $H$, the $U(1)_{\mathrm{PQ}}$ transformation becomes
\begin{align}
  H \to e^{-i\alpha}He^{i\alpha \sigma_3 \cos2\beta}.
\end{align}


\subsection{Definition of unbroken $U(1)_{\mathrm{EM}}$ group}
Unlike in the vacuum, in the presence of a soliton background, 
the definition of the unbroken $U(1)_{\mathrm{EM}}$ generator is non-trivial.
In this paper,
it is defined as \footnote{
The vector $n^a$ corresponds to $\tilde{n}^a$ in Ref.~\cite{Eto:2020opf}.
}
\begin{equation}
 \hat{Q} H \equiv - n^a \f{\sigma_a}{2}H - H \f{\sigma_3}{2}, \label{010004_1Oct20}
\end{equation}
where 
\begin{equation}
 n^a \equiv \f{\sum_{i=1,2}|H_i|^2 n_i^a}{C},
\end{equation}
\begin{equation}
n_1^a \equiv \f{H_1 ^\dagger \sigma^a H_1}{|H_1|^2}, \h{2em} n_2^a \equiv \f{H_2 ^\dagger \sigma^a H_2}{|H_2|^2}.
\end{equation}
The positive normalization factor $C$ is determined to satisfy $n^a n^a =1$.
Correspondingly, the $U(1)_Z$ subgroup in the $SU(2)_W \times U(1)_Y$ group is defined as
\begin{equation}
 \hat{T}_Z H \equiv - n^a \f{\sigma_a}{2} H - \sin ^2 \theta_W \hat{Q}H . 
\end{equation}
Also, the $U(1)_Z$ and $U(1)_\mathrm{EM}$ gauge fields are defined as
\begin{equation}
Z_\mu \equiv - n^a W_\mu ^a \cos\theta_W - Y_\mu \sin\theta_W,\label{012344_6Oct20}
\end{equation}
\begin{equation}
A_\mu \equiv - n^a W_\mu ^a \sin\theta_W + Y_\mu \cos\theta_W.\label{012356_6Oct20}
\end{equation}
In addition, the charged components of $SU(2)_W$ gauge group is defined as orthogonal components to $n^a \sigma_a$.
In the vacuum,
the Higgs field takes a constant VEV $\langle H \rangle = \mathrm{diag} (v_1,v_2)$,
and $n^a \sigma_a = - \sigma_3$.
The above definitions reduce to the conventional ones.
The VEV is invariant under $U(1)_{\mathrm{EM}}$,
\begin{equation}
 \hat{Q}\langle H \rangle =0,
\end{equation}
which means that the $U(1)_{\mathrm{EM}}$ symmetry is not spontaneously broken in the vacuum.

It may be useful to rewrite the above expressions for the two doublets $H_1 $ and $H_2$,
\begin{align}
  \hat{Q} H_i = &\left( - n^a \f{\sigma_a}{2} + \f{1}{2} \bm{1}\right) H_i, \label{010017_1Oct20}
 \\
 \hat{T}_Z H_i = &\left(- n^a \f{\sigma_a}{2} - \sin ^2 \theta_W \hat{Q}\right) H_i ,
\end{align}
for $i=1,2$.

\section{Electroweak axion strings}
\label{sec:electroweakaxionstrings}
Similarly to other axion models, the DFSZ axion model provides a vortex string solution 
known as the axion string corresponding to the breaking of $U(1)_{\mathrm{PQ}}$.
On the other hand, after the electroweak phase transition, 
the axion string can contain flux tubes of the $SU(2)_W \times U(1)_Y$ gauge fields like the Abrikosov-Nielsen-Olesen vortex \cite{Abrikosov:1956sx,Nielsen:1973cs}
since the two Higgs doublets also acquire the VEVs.
We call such vortex strings the electroweak axion strings.
In this section, we show that there are (at least) three types of the electroweak axion strings in the DFSZ model.
Interestingly, some of them have similar properties to those of (non-Abelian) vortices in two Higgs doublet models,
in which there is a global symmetry for a relative rotation of the two Higgs doublets.
In particular, some part of our argument in this section refers to that in Refs.~\cite{Eto:2018hhg,Eto:2018tnk}.


\subsection{Axion string in DFSZ model}
We first review the conventional axion string in this subsection.
Let us consider a case that the $U(1)_{\mathrm{PQ}}$ symmetry is spontaneously broken by $\langle S \rangle \neq 0$ 
but the electroweak symmetry remains,
$\langle H_1 \rangle = \langle H_2 \rangle =(0,0)^{\mathrm{T}}$.
This situation realized in the early universe when the temparature $T_{th}$ satisfies $v_{EW}\ll T_{th} \ll v_s$.
In this case, as is well-known, a vortex-string configuration associated with the global $U(1)_{\mathrm{PQ}}$ symmetry exists as a solution to the equation of motion (EOM),
which is called the axion string in the literature.
The configuration located on the $z$-axis is described by the following ansatz
\begin{equation}
 S= v_s e^{i \theta} \phi(r) ,\hspace{ 2em} H_1 = H_2 = \begin{pmatrix}0 \\ 0 \end{pmatrix},\label{041816_30Sep20} 
\end{equation}
where $r$ and $\theta$ are the distance from the $z$-axis and the rotational angle, respectively.
Namely, $x+iy = r e^{i\theta}$.
The profile function $\phi(r)$ satisfies
the boundary conditions
\begin{equation}
 \phi(0)=0, \hspace{2em} \phi( \infty) =1.
\end{equation}
The detailed form of $\phi(r)$ is determined by solving the EOM.
This string has a winding number associated with the $U(1)_{\mathrm{PQ}}$ symmetry, 
and hence is topologically stable.
The $U(1)_\mathrm{PQ}$ symmetry is restored on the string core because of $\phi(0)=0$.

It is known that such strings are necessarily produced 
during the phase transition of the $U(1)_{\mathrm{PQ}}$ symmetry breaking
by the Kibble-Zurek mechanism.
In the viewpoint of phenomenology,
one of the important aspects of cosmic strings is the interaction between a pair of the cosmic strings having the same topological charge.
For axion strings, the interaction is dominated by exchange of massless axion particles, 
resulting in the long-range repulsive force.
The potential of the interaction $V_\mathrm{st.}$ is approximately given as $V_\mathrm{st.} \sim - v_s^2\log R$ with $R$ being the distance between the pair.
Due to the repulsive interaction, 
a pair of the strings reconnects with probability of the order of unity when they collide to each other
and does not form a bound state of the strings (such as the Y-junction \cite{Bettencourt:1996qe,Bettencourt:1994kc,Copeland:2006eh,Copeland:2006if,Salmi:2007ah,Bevis:2008hg,Bevis:2009az,Hiramatsu:2013yxa,Hiramatsu:2013tga}).
As a result, the strings form a stationary network
whose typical length scale remains to be the Hubble horizon scale (scaling regime).
Such a scale-invariant evolution of the network prevents the energy density of the strings from dominating that of the universe,
and thus axion models producing the axion strings are cosmologically viable as far as concerning the strings.

\subsection{Vortex string with $Z$-flux (type-A string)}

Next, we discuss the electroweak axion strings.
Let us consider a string configuration after the electroweak phase transition $T_{th} \lesssim v_\mathrm{EW}$.
The two doublets also acquire the VEVs
and their phases must also wind because they also have the $U(1)_\mathrm{PQ}$ charges, 
otherwise divergent energy arises from $V_\mathrm{mix}$.%
\footnote{If the two doublets had no windings, the mixing term provides $\kappa v_s^2 v_1 v_2 \cos2\theta $ at large distances, which means a divergent potential energy after the spatial integration.}
From the single-valuedness of the doublets, 
the string configuration has the form 
\begin{equation}
 \begin{cases}
   S= v_s e^{i \theta} \phi(r) \\
  H_1 = v_1 e^{ i\theta}\begin{pmatrix}0 \\ f(r)\end{pmatrix} \\
  H_2 = v_2 e^{-i\theta}\begin{pmatrix}0 \\ h(r) \end{pmatrix}
 \end{cases}\label{153047_3Oct20},
\end{equation}
\begin{equation}
 Z_i = \frac{2\cos 2\beta }{g_Z}\frac{ \epsilon_{ij} x_j}{r^2 }
 (1- z(r)) ,
 \label{231721_10Sep20}
\end{equation}
and $W_i^{\pm}= A_i=0$.
$g_{Z}$ is the coupling of $Z$-boson given by $g_Z = \sqrt{g^2 + g'^2}$.
$\epsilon_{ij}$ is the anti-symmetric tensor satisfying $\epsilon_{12} = -\epsilon_{21} = 1$.
We call this string configuration the type-A electroweak axion string.

The last two configurations in Eq.~\eqref{153047_3Oct20} are equivalent to 
\begin{equation}
 H= e^{- i\theta } \begin{pmatrix} v_1 f(r)  & 0 \\ 0 & v_2 h(r) \end{pmatrix}.
\end{equation}
The profile functions $f(r),h(r)$ satisfy the same boundary conditions as that of $\phi(r)$, i.e.,
\begin{equation}
 \phi(0)=f(0)=h(0)=0, \hspace{2em} \phi(\infty)=f( \infty)= h(\infty) =1.\label{231852_15Oct20}
\end{equation}
The profile function for the gauge field $z(r)$ should satisfy the boundary conditions
\begin{align}
 z(0) = 1,
 \quad
 z(\infty) =0.
 \label{153351_3Oct20}
\end{align}
Eq.~\eqref{231852_15Oct20} indicates that the electroweak symmetry is restored inside the string as well as $U(1)_\mathrm{PQ}$.

Noting that the two doublets have $U(1)_{\mathrm{PQ}}$ charges as given in Eq.~\eqref{224951_10Sep20},
it is convenient to decompose the winding phases as
\begin{align}
  H_1 & = v_1 e^{2i \theta s_\beta^2 } e^{-i\theta \sigma_3 c_{2\beta}}\begin{pmatrix}
  0 \\
  f(r)
\end{pmatrix},
  \label{030036_11Sep20} \\
  H_2 & = v_2 e^{-2i \theta c_\beta^2 }  e^{-i\theta \sigma_3 c_{2\beta}}
\begin{pmatrix}
 0 \\
 h(r)
\end{pmatrix},\label{030041_11Sep20}
\end{align}
with $ c_X \equiv \cos(X)$ and $ s_X \equiv \sin(X)$.
Eqs.~\eqref{030036_11Sep20} and \eqref{030041_11Sep20} mean that the configurations of the doublets have the winding number unity for the global $U(1)_{\mathrm{PQ}}$ symmetry and the fractional winding number $- \cos 2 \beta$ for the $U(1)_Z$ subgroup of the $SU(2)_W \times U(1)_Y$ gauge symmetry.
Therefore, the gradient energy from the $U(1)_Z$ windings is canceled by the $Z$ gauge field \eqref{231721_10Sep20} with Eq.~\eqref{153351_3Oct20} at large distances $r\to \infty$.
It follows from Eq.~\eqref{231721_10Sep20} that the string configuration has the $Z$-flux,
\begin{equation}
 \Phi_Z = \oint_{r= \infty} dx_i~ Z_i= \frac{ -4 \pi \cos 2\beta}{g_Z},
\end{equation}
which is fractionally quantized because of the fractional winding number.

\begin{figure}[t]
\centering
\includegraphics[width=0.5\textwidth]{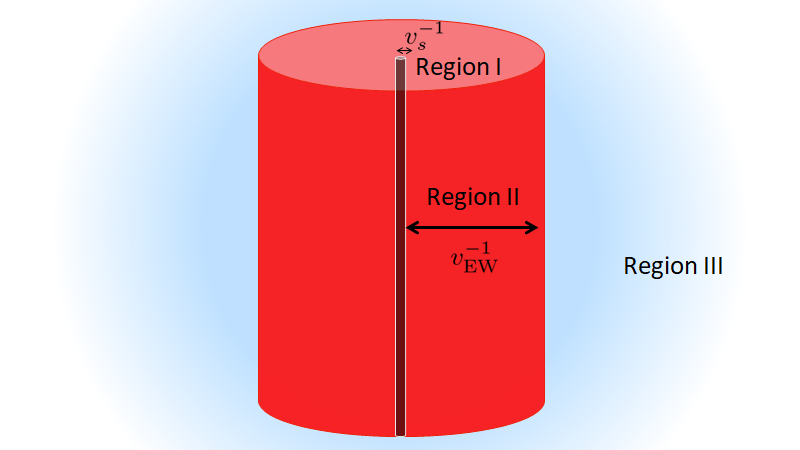}
\caption{
Schematic picture of the energy density profile of the electroweak axion strings (type-A, B and C).
The energy density consists of three parts.
There is a one-dimensional thin object consisting of the radial component of the complex scalar $\phi(r)$ in Region I: $r \lesssim v_s^{-1}$.
We call this region the core of the string.
In Region II: $v_s^{-1} \lesssim r \lesssim (v_\mathrm{EW})^{-1}$ (red region),
the energy density is dominated by the Higgs fields and the electroweak gauge fields.
This region is much fatter than Region I\@.
In addition, there is the fattest part made from the gradient energy of the axion, leading to the logarithmically divergent tension.
This is denoted by Region III: $r \gtrsim (v_\mathrm{EW})^{-1}$ (blue cloud).
Note that the type-A string with $\tan \beta =1$ is a special case
since it has only the global winding and its energy density does not have the Region II\@.
}
\label{214207_5Oct20}
\end{figure}

We discuss a qualitative property of the profile functions.
The typical length scale for $\phi(r)$ is $v_s^{-1}$
while those of $f(r),h(r)$ and $z(r)$ are given as $(v_{\mathrm{EW}})^{-1}$.
Thus the string is a ``multi-scale solution''.
Fig.~\ref{214207_5Oct20} shows a schematic picture of the energy density profile of the electroweak axion strings.
The energy density has three structures.
One is that from the radial component of the complex scalar, $\phi(r)$, whose typical scale is  $v_s^{-1}$.
This part looks as a thin object (Region I: $r \lesssim v_s^{-1}$).
Another is from those of the two Higgs doublets and the gauge fields, whose typical scale is the EW-scale $(v_\mathrm{EW})^{-1}$.
This region, which we call Region II, is much fatter than Region I and is shown as the red region.
The third part is the fattest part from the gradient energy of the axion.
This is denoted by Region III: $r \gtrsim (v_\mathrm{EW})^{-1}$ (blue cloud in the figure).
When one calculates the tension of the string (energy per unit length) by integrating the energy density on the $xy$ plane,
the third part leads to the log-divergent tension $\sim 2\pi v_s^2 \log L$ 
where $L$ is the IR-cutoff and is usually taken as the distance between neighbor two strings.
The coefficient of the log divergence is the same as that of the conventional axion string shown in the last subsection
because it depends only on the winding number of the global $U(1)_{\mathrm{PQ}}$ symmetry.

Let us obtain the profile functions and calculate the string tension for the type-A string in a numerical way.
For simplicity, we take $m_2^2=\alpha_4=0$ and $\kappa_{1S}=\kappa_{2S}$,
leading to $\tan \beta=1$ (see the stationary condition \eqref{143600_6Oct20}).
The VEVs are denoted as $v_1=v_2 \equiv v$.
We should note that in this case, the Higgs doublets do not have winding number for the $U(1)_Z$ gauge subgroup,
and thus the right hand side of Eq.~\eqref{231721_10Sep20} vanishes.
The $Z$-flux is constantly zero.
After substituting the ansatz, the energy density is given by
\begin{align}
 \mathcal{E} \equiv & ~ |\partial_i S|^2 + \mathrm{Tr} |D _i H|^2 
 + \frac{1}{4} (W_{ij}^a)^2 + \frac{1}{4} (Y_{ij})^2 + V(H,S)\label{172833_3Oct20} 
 \\
 =& ~\frac{v^2}{r^2} \left[r^2
 \left(f'(r)^2+h'(r)^2\right)+f(r)^2+h(r)^2\right]\n
 \\
 &+v^2 \left[-m_1^2 (f(r)^2 + h(r)^2)+ 2 \alpha_2 v^2 f(r)^2 h(r)^2
 +v^2\alpha_{123} (f(r)^4 + h(r)^4) \right] \n \\
 & +v^2 v_s^2
 \left[2 \kappa f(r) h(r) \phi (r)^2+\kappa_{1S} (f(r)^2+h(r)^2) \phi(r)^2 \right] \n
 \\
 & +v_s^2 \left(- m_S^2 \phi (r)^2+ \lambda_S v_S^2 \phi(r)^4 \right)
 +\frac{v_s^2}{r^2} \left(r^2 \phi '(r)^2+\phi (r)^2\right),
\end{align}
with $\alpha_{123}\equiv \alpha_1 + \alpha_2 + \alpha_3$
and $'$ denoting the derivative with respect to $r$.

\begin{figure}[tbp]
 \centering
\includegraphics[width=0.47\textwidth]{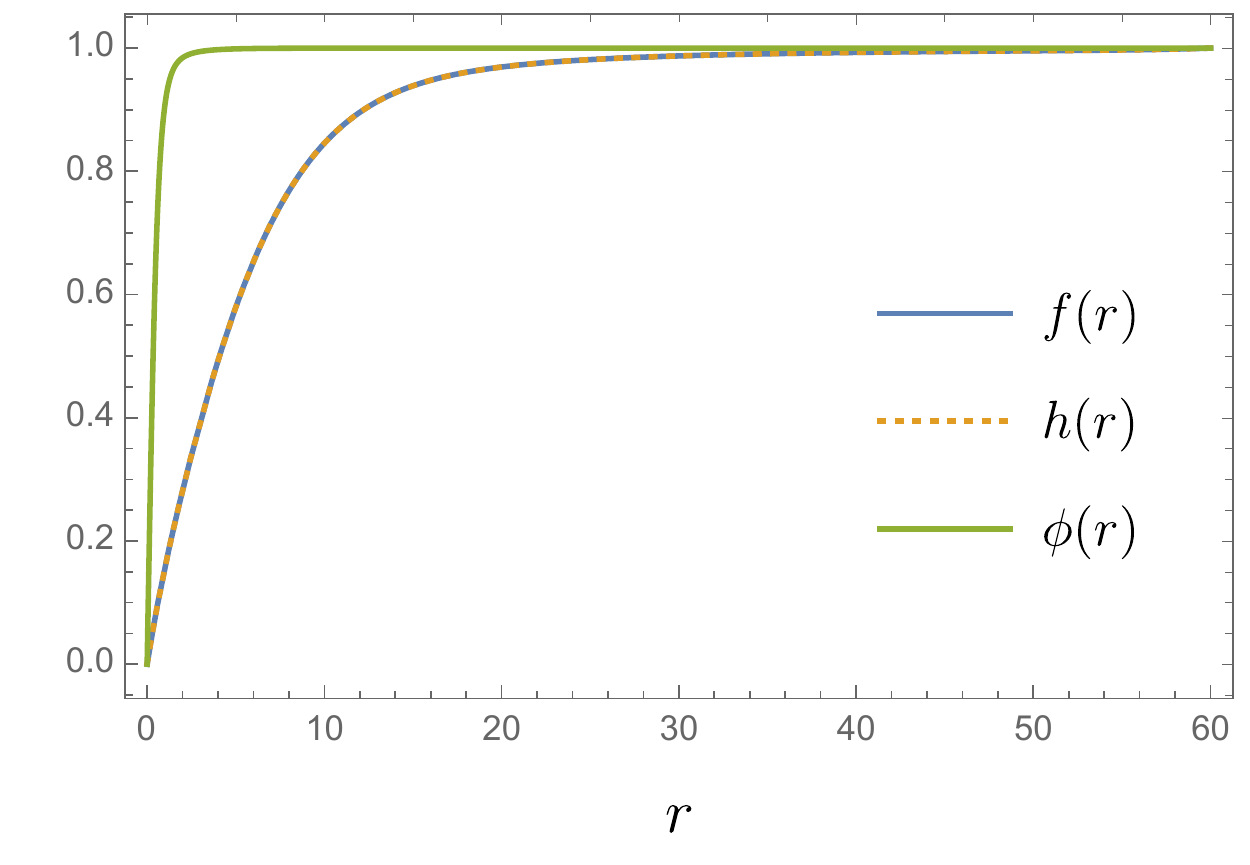} \hspace{0.5em}
\includegraphics[width=0.5\textwidth]{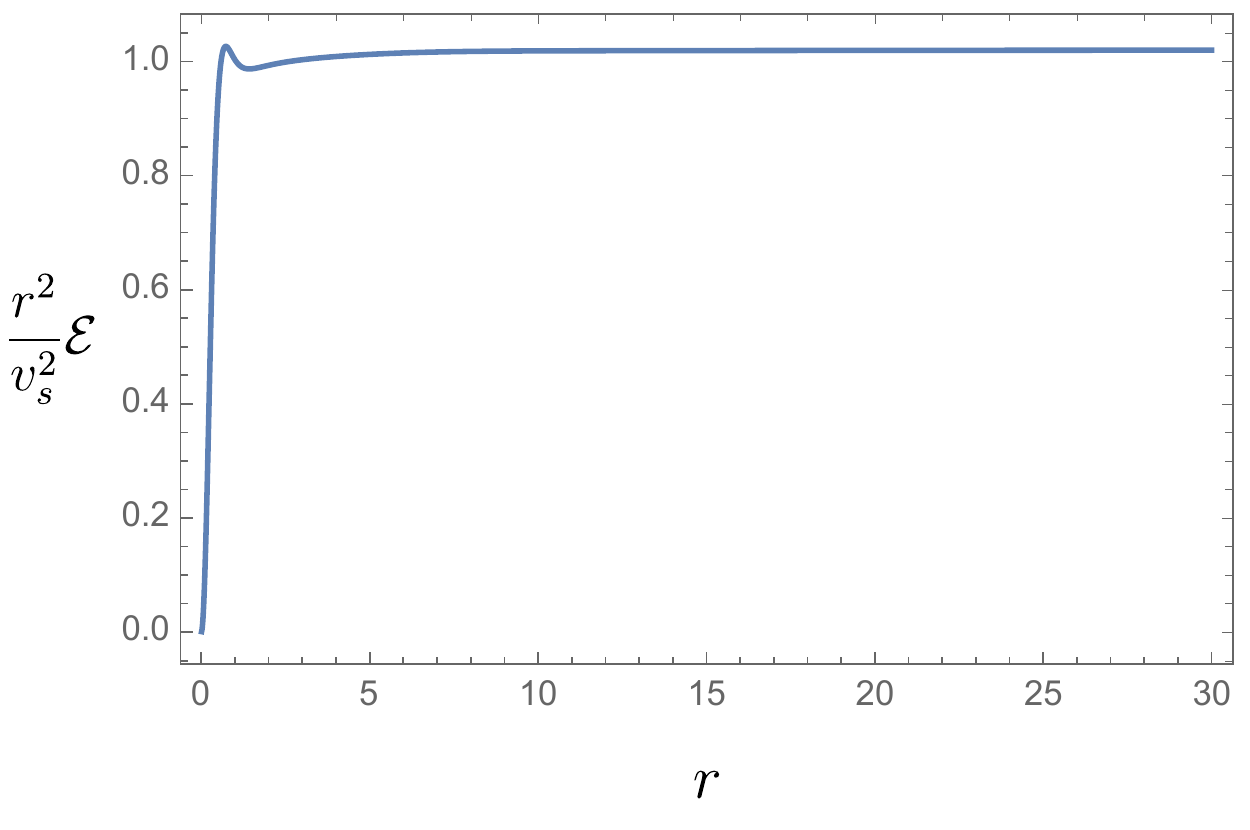}
\caption{
Numerical solution for the type-A string.
We take benchmark parameters as Eq.~\eqref{052110_30Sep20} and $v_s=10\, v$.
Also we adopt a length unit as $v_s^{-1}=0.5$.
(left): Plots of profile functions. Note $f(r)$ (blue line) is equal to $h(r)$ (dotted orange line) everywhere.
$\phi(r)$ increases as $\phi \sim v_s r$ for $r \sim 0$ while $f =h \sim v r$.
All of them approach to unity for $r\to \infty$.
(right): Plot of energy density $\mathcal{E}$ (Eq.~\eqref{172833_3Oct20}) divided by $v_s^2/r^2$.
Clearly, the energy density has a polynomial tail like $r^{-2}$, instead of an exponential one.
This leads to the logarithmically divergent tension.
The integrated value of the tension over $0 \leq r \leq 120 \,v_s^{-1}$ is $140.321$.
}
\label{044741_30Sep20}
\end{figure}

The EOMs are obtained as
\begin{align}
 &f''(r)+\frac{f'(r)}{r}
 -\frac{f(r)}{r^2}
 \nonumber\\
 &- \Bigl( 2 \alpha_{123}\, v^2 f(r)^2 +2 \alpha_2 v^2 h(r)^2 +  \kappa_{1S} v_s^2 \phi(r)^2 -m_1^2 \Bigr) f(r)
 - \kappa v_s^2 h(r) \phi(r)^2 =0, \\
 & h''(r)+ \frac{h'(r)}{r} - \frac{h(r)}{r^2}
 \nonumber\\
 &- \Bigl( 2 \alpha_{123}\, v^2 h(r)^2 + 2 \alpha_2 v^2 f(r)^2 + \kappa_{1S} v_s^2 \phi(r)^2 -m_1^2 \Bigr) h(r)
 -\kappa v_s^2 f(r) \phi(r)^2 =0,
 \\
 &\phi''(r) + \frac{\phi'(r)}{r} - \frac{\phi(r)}{r^2}
 \nonumber\\
 &-\Bigl( 2 \lambda_S v_s^2 \phi(r)^2 + 2 \kappa_{1S} v^2 (f(r)^2 + h(r)^2) + 2 \kappa v^2 f(r) h(r) - m_S^2 \Bigr) \phi (r) =0.
\end{align}
We adopt the so-called relaxation method to solve the EOMs.
As a benchmark case, we take the parameters as
\begin{equation}
\alpha_1=1,\h{1em} \alpha_2=-0.3348, \h{1em} \alpha_3=0, \h{1em}\lambda_S=1, \h{1em}\kappa=-2 \left(\f{v}{v_s}\right)^2, \h{1em} \kappa_{1S}=0.4 , \label{052110_30Sep20}
\end{equation}
such that the lightest scalar mass $m_{h_1}^2$ reproduces the SM Higgs mass $(125 \text{ GeV})^2$.
In addition, we set the VEV for $S$ as $v_s=10 \, v$.
Although this is too small and not viable in the phenomenological viewpoint, 
it does not matter because the qualitative picture of the type-A string does not change.
If one takes them more hierarchical, huge numerical costs arise in the calculation.
The obtained numerical solutions are shown in Fig.~\ref{044741_30Sep20}. 
In the left panel, $f(r)$ (blue line) is equal to $h(r)$ (dotted orange line) everywhere.
$\phi(r)$ increases as $\phi \sim v_s r$ for $r \sim 0$ while $f =h \sim v r$.
All of them approach to unity for $r\to \infty$.
The right panel shows the energy density $\mathcal{E}$ (Eq.~\eqref{172833_3Oct20}) divided by $v_s^2/r^2$.
The divided value approaches to unity, which means that 
the energy density has a polynomial tail like $r^{-2}$, instead of an exponential one.
This leads to the logarithmically divergent tension.

\subsection{Type-B string with $Z$-flux}

\begin{figure}[tbp]
 \centering
\includegraphics[width=0.55\textwidth]{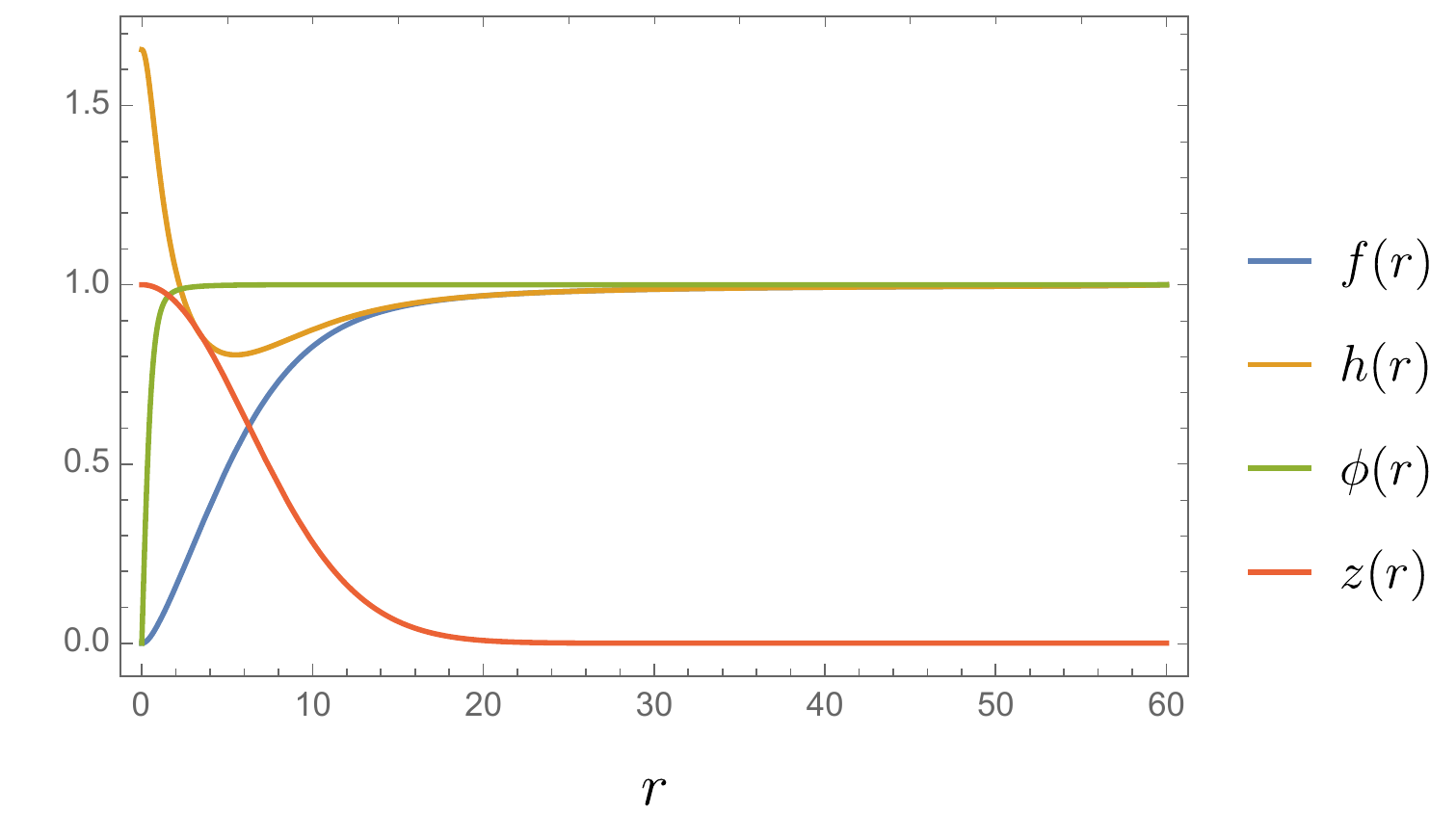} \hspace{0.em}
\includegraphics[width=0.43\textwidth]{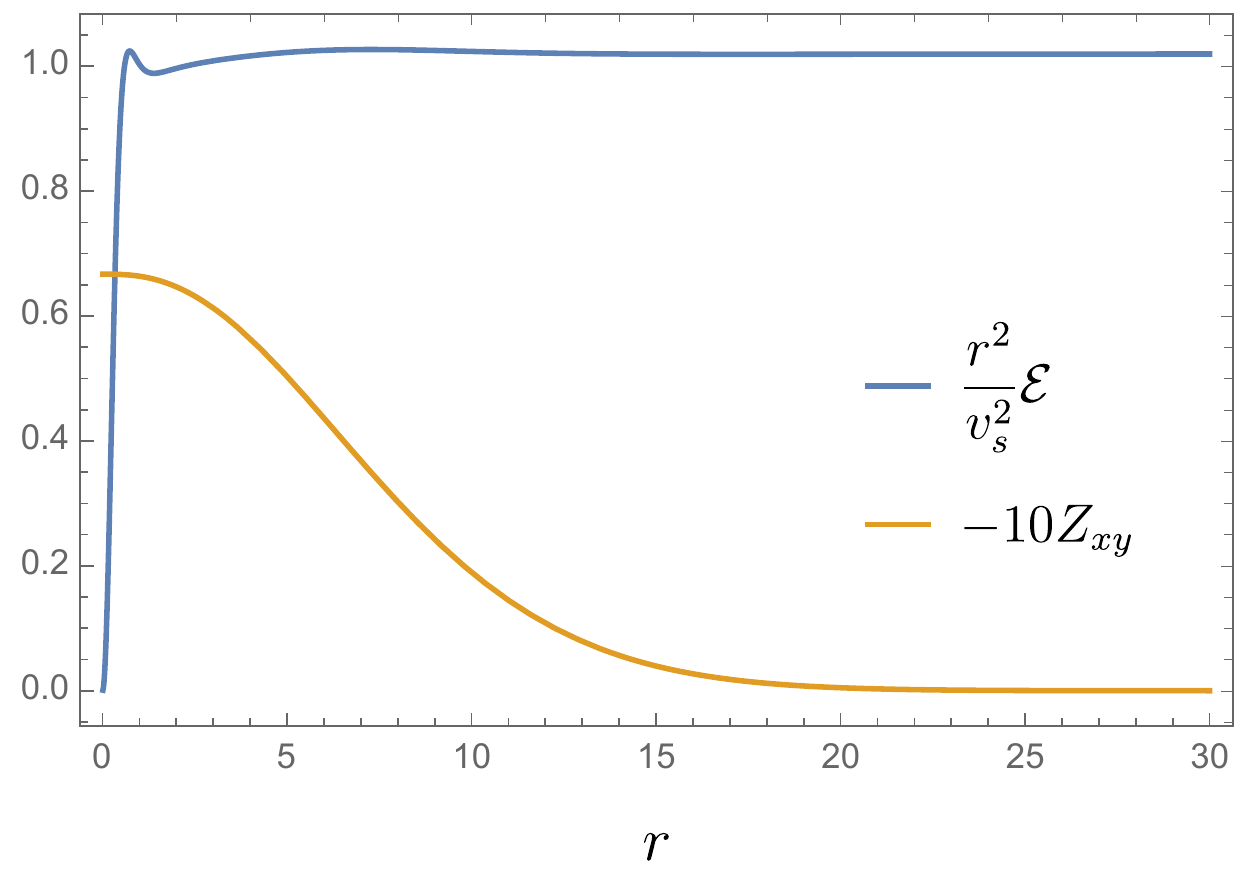}
\caption{
Numerical solution for the type-B string.
We take the same parameters as ones in Fig.~\ref{044741_30Sep20}.
Also we adopt a length unit as $v_s^{-1}=0.5$.
(left): Plots of profile functions. 
$\phi(r)$ increases as $\phi \sim v_s r$ for $r \sim 0$
while $f(r)$ behaves as a quadratic function with respect to $r$.
The three profile functions $f,h$ and $\phi$ approach to unity for $r\to \infty$.
The profile function of the $Z$ field, $z(r)$, approaches to zero as $r \to \infty$ starting from unity at $r=0$.
(right): Plots of energy density \eqref{193824_3Oct20} divided by $v_s^2/r^2$ 
and the $Z$-flux density multiplied by $-10$.
The energy density has a polynomial tail like $r^{-2}$ like the type-A string.
The tension $T$ integrated over $0 \leq r \leq 120 \, v_s^{-1}$ is $140.524$.
The $Z$-flux density decays exponentially for $r\to \infty$ like the usual Abrikosov-Nielsen-Olesen vortex.
The total value of the $Z$-flux is calculated to $-16.9539$, which is consistent with Eq.~\eqref{051939_30Sep20}.
}
\label{045818_30Sep20}
\end{figure}

We have shown that
the phases of the two Higgs doublets must wind after the electroweak phase transition $T_{th} \lesssim v_\mathrm{EW}$.
Actually, there is another type of vortex string that is consistent with the single-valuedness and the potential minimum.
This is obtained by giving an \textit{additional winding} in the $U(1)_Z$ gauge orbit to the type-A string.
We call this string the type-B electroweak axion string.
The ansatz describing the string is given as
\begin{equation}
 \begin{cases}
   S= v_s e^{i \theta} \phi(r) \\
  H_1 = v_1 e^{ 2i\theta}\begin{pmatrix}0 \\ f(r)\end{pmatrix} \\
  H_2 = v_2 \begin{pmatrix}0 \\ h(r) \end{pmatrix}
 \end{cases}\label{194425_3Oct20}
\end{equation}
\begin{equation}
 Z_i = \frac{ 4\cos^ 2\beta }{g_Z}\frac{ \epsilon_{ij} x_j}{r^2 }
 (1-z(r)).
 \label{174611_3Oct20}
\end{equation}
The last two configurations in Eq.~\eqref{194425_3Oct20} are equivalent to 
\begin{equation}
 H= e^{- i\theta }e^{- i\theta \sigma_3}
 \begin{pmatrix}
 v_1 f(r)  & 0 \\ 
 0 & v_2 h(r)
 \end{pmatrix}.
\end{equation}
The profile functions $f(r)$ and $\phi(r)$ satisfy the same boundary conditions,
\begin{equation}
\phi(0)= f(0)=0, \hspace{1em} \phi(\infty)= f( \infty) =1\label{232306_30Sep20},
\end{equation}
but $h(r)$ should satisfy the following boundary conditions:
\begin{equation}
 \partial_r h|_{r=0}=0, \hspace{1em} h(\infty)= 1.
 \label{eq:typeBhbc}
\end{equation}
The profile function for the gauge field $z(r)$ satisfies 
\begin{align}
 z(0) =1 ,
 \quad
 z(\infty) =0.
\end{align}
Note that $h(r)$ is not fixed to zero on the center of the string because it does not have a winding phase.
Therefore, the electroweak symmetry is \textit{not} restored inside the string core (see Fig.~\ref{045818_30Sep20}.)

Again, we decompose the winding phases as
\begin{align}
  H_1 & = v_1 e^{2i \theta s_\beta^2 } e^{-2i\theta \sigma_3 c_{\beta}^2}\begin{pmatrix}0 \\ f(r)\end{pmatrix}, \\
  H_2 & = v_2 e^{-2i \theta c_\beta^2 }  e^{-2i\theta \sigma_3 c_{\beta}^2}\begin{pmatrix}0 \\ h(r) \end{pmatrix},
\end{align}
which mean that the configurations of the doublets have the winding number unity for the global $U(1)_{\mathrm{PQ}}$ symmetry and the fractional winding number $ -2 \cos^2 \beta$ for the $U(1)_Z$ subgroup.
The difference between the winding number of the type-A and type-B strings is unity.
Similarly to the previous case, the $Z$ gauge field \eqref{174611_3Oct20} cancels the gradient energy from the $U(1)_Z$ windings.
It follows from Eq.~\eqref{231721_10Sep20} that the string configuration has the $Z$-flux,
which is calculated as
\begin{equation}
 \Phi_Z = \oint_{r= \infty} dx_i~ Z_i= \frac{ -8 \pi \cos^ 2\beta}{g_Z}.
 \label{051939_30Sep20}
\end{equation}
Interestingly, this string configuration is quite similar to the topologically stable $Z$-string (topological vortex with the $Z$-flux) in 2HDM \cite{Dvali:1993sg,Dvali:1994qf,Eto:2018hhg,Eto:2018tnk}.
In addition to the winding number of $U(1)_Z$,
the difference from the 2HDM is the existence of
the singlet complex scalar carrying the $U(1)_{\mathrm{PQ}}$ charge,
which is replaced by the relative phase rotation of
the two doublets in the 2HDM.

Let us obtain the profile functions and calculate the string tension for the type-B string in a numerical way.
Again, we take $m_2^2 = \alpha_4=0$ and $\kappa_{1S}=\kappa_{2S}$,
leading to $\tan \beta=1$.
The VEVs are denoted as $v_1=v_2 \equiv v$.
Unlike the type-A string, 
the Higgs doublets have winding numbers for the $U(1)_Z$ gauge subgroup even for $\tan \beta=1$.
Consequently, the $Z$-flux is non-zero and confined inside the string.
The energy density is
\begin{align}
 \mathcal{E}
 = & \frac{v^2}{r^2} \left[r^2
   \left(f'(r)^2+h'(r)^2\right)+f(r)^2(1+z(r))^2+h(r)^2(1-z(r))^2\right]\n\label{193824_3Oct20} 
 \\
 &+v^2 \left[-m_1^2 (f(r)^2 + h(r)^2)+ 2 \alpha_2 v^2 f(r)^2 h(r)^2
 +v^2\alpha_{123} (f(r)^4 + h(r)^4) \right] \n \\
 & +v^2 v_s^2
   \left[2 \kappa f(r) h(r) \phi (r)^2+\kappa_{1S} (f(r)^2+h(r)^2) \phi(r)^2\right] \n
 \\
 & +v_s^2 \left(- m_S^2 \phi (r)^2+ \lambda_S v_S^2 \phi(r)^4 \right)+\frac{v_s^2}{r^2} \left(r^2 \phi '(r)^2+\phi (r)^2\right) 
+2 \f{z'(r)^2}{g_Z^2 r^2}
\end{align}
Then, the EOMs are obtained as
\begin{align}
 &f''(r) + \frac{f'(r)}{r} - \frac{(1+z(r))^2}{r^2} f(r)
 \nonumber\\
 &- \Bigl( 2 \alpha_{123}\, v^2 f(r)^2 + 2\alpha_2 v^2 h(r)^2 + \kappa_{1S}v_s^2 \phi(r)^2 
 -m_1^2 \Bigr) f(r)- \kappa v_s^2 h(r) \phi(r)^2 =0,
\end{align}
\begin{align}
 &h''(r) + \frac{h'(r)}{r}- \frac{(-1 + z(r))^2}{r^2} h(r)
 \nonumber\\
 &- \Bigl( 2\alpha_{123} \, v^2 h(r)^2+ 2 \alpha_2 v^2 f(r)^2 + \kappa_{1S} v_s^2 \phi(r)^2
 - m_1^2 \Bigr) h(r) - \kappa v_s^2 f(r) \phi(r)^2 =0,
 \\
 &\phi''(r) + \frac{\phi'(r)}{r} - \frac{\phi(r)}{r^2}
 \nonumber\\
 &- \Bigl( 2 \lambda_S v_s^2 \phi(r)^2 + \kappa_{1S} v^2 (f(r)^2 + h(r)^2)+ 2\kappa v^2 f(r) h(r) -m_S^2 \Bigr) \phi(r)=0 ,
 \\
 &z''(r)- \frac{z'(r)}{r} - \frac{g_Z^2 v^2 }{2} f(r)^2 ( 1 + z(r)) - \frac{g_Z^2 v^2}{2} h(r)^2 ( -1 + z(r) ) =0.
\end{align}
We take the same parameter choice as Eq.~\eqref{052110_30Sep20}.
In addition, we set the VEV for $S$ as $v_s=10 \, v$.
The obtained numerical solutions are shown in Fig.~\ref{045818_30Sep20}.
In the left panel, $f(r)$ ($\phi(r)$) behaves like $v_s r$ ($v r^2$) at the origin
because the phases of $\phi(r)$ and $f(r)$ wind once and twice, respectively.
$h(r)$ does not start from zero at the origin
due to the Neumann condition at the origin.
All of the scalar profile functions approach to unity for $r\to \infty$.
The profile function for the gauge field $z(r)$ approaches to zero starting from unity.
The right panel shows the energy density $\mathcal{E}$ (Eq.~\eqref{172833_3Oct20}) divided by $v_s^2/r^2$ 
and the $Z$-flux density multiplied by $-10$.
Similarly to the previous case,
the type-A string,
the energy density has a polynomial tail like $r^{-2}$, instead of an exponential one.
This leads to the logarithmically divergent tension.
The schematic picture of the energy density profile is the same as Fig.~\ref{214207_5Oct20}.
On the other hand, the $Z$-flux has an exponential tail.

In the above ansatz for the type-B string, only the one doublet $H_1$ has the winding number.
There may be also an alternative string in which only $H_2$ has the winding phase.
Roughly speaking, they are related by exchange of the two doublets $H_1$ and $H_2$.
Since the property is similar to the former, 
we do not study the latter one in this paper and also categorize the latter one as the type-B string.

\subsection{Type-C string with $W$-flux}

Finally, we consider the third type of the vortex string, called the type-C electroweak axion string.
While the type-B string has been obtained by performing an additional rotation of $U(1)_Z$ on the type-A one,
the type-C string has a winding in the $U(1)_{W^1}$ subgroup of the $SU(2)_W \times U(1)_Y$ symmetry.
The ansatz for the scalar fields is given as
\begin{equation}
 \begin{cases}
   S= v_s e^{i \theta} \phi(r) \\
  H_1 = \frac{1}{2} v_1 e^{ i\theta} 
\begin{pmatrix} f(r)e^{i\theta} - h(r)e^{-i\theta} \\ f (r)e^{i\theta} + h(r)e^{-i\theta} \end{pmatrix} \\
  H_2 = \frac{1}{2} v_2 e^{ - i\theta}
\begin{pmatrix} h(r)e^{i\theta} - f(r)e^{-i\theta} \\ h(r)e^{i\theta} + f(r)e^{-i\theta} \end{pmatrix} \\
 \end{cases}\label{235219_22Sep20}
\end{equation}
The profile functions $f(r),h(r), \phi(r)$ satisfy similar boundary conditions to those of the type-B string,
\eqref{232306_30Sep20} and \eqref{eq:typeBhbc},
i.e.,
\begin{equation}
 f(0)=\phi(0)=0, \hspace{1em} \partial_r h|_{r=0}=0, \hspace{1em} f( \infty)= h(\infty) =\phi(\infty)=1.\label{235011_30Sep20}
\end{equation}
Due to the non-zero value of $h(0)$, the electroweak symmetry is not restored inside the string as the type-B string.

Unlike the type-A and type-B strings,
in the background of the type-C string,
the $U(1)_{\mathrm{EM}}$ and $U(1)_Z$ generators
depend on the positions.
As explained in Sec.~\ref{005936_1Oct20}, 
the $U(1)_{\mathrm{EM}}$ generator is defined by Eq.~\eqref{010004_1Oct20} or Eq.~\eqref{010017_1Oct20}.
Substituting the ansatz Eq.~\eqref{235219_22Sep20}, we obtain 
\begin{align}
 n_1^a =& \f{H_1 ^\dagger \sigma^a H_1}{|H_1|^2}  =  \frac{2}{f^2+h^2}\left((f^2-h^2)/2, -fh \sin 2\theta, -fh \cos 2\theta\right),
 \\
 n_2^a =& \f{H_2 ^\dagger \sigma^a H_2}{|H_2|^2}  = \frac{2}{f^2+h^2}\left((h^2-f^2)/2, -fh \sin 2\theta, -fh \cos 2\theta\right),
\end{align}
and
\begin{equation}
 n^a \f{\sigma_a}{2}= - \f{\sigma_2}{2} \sin 2\theta - \f{\sigma_3}{2} \cos 2\theta.
\end{equation}
Then the $U(1)_{\mathrm{EM}}$ and $U(1)_Z$ generators are given by
\begin{equation}
 \hat{Q} H_i = \left( \f{\sigma_2}{2} \sin 2\theta + \f{\sigma_3}{2} \cos 2\theta + \f{1}{2} \bm{1} \right)H_i,
\end{equation}
\begin{equation}
 \hat{T}_Z H_i =  \left(\sigma_2 \sin 2\theta + \sigma_3 \cos 2\theta - \sin^2\theta_W \hat{Q}\right) H_i,
\end{equation}
which depend on $\theta$.

Let us see the asymptotic behaviors of the two doublets at large distances $r\to \infty$.
From Eq.~\eqref{235011_30Sep20}, we obtain
\begin{align}
 H_1 &\sim  v_1 e^{ i\theta} 
 \begin{pmatrix}
 i \sin \theta \\
 \cos \theta
 \end{pmatrix}
 = v_1 e^{ 2i\theta s_\beta ^2} e^{ -i\theta c_{2\beta}\sigma_Z}
 e^{ i\theta \sigma_1} 
 \begin{pmatrix}
 0 \\
 1
 \end{pmatrix},
 \\
 H_2 &\sim  v_2 e^{-i\theta} 
 \begin{pmatrix}
 i \sin \theta \\
 \cos \theta
 \end{pmatrix}
 = v_2 e^{ -2i\theta c_\beta ^2} e^{ -i\theta c_{2\beta}\sigma_Z}
 e^{ i\theta \sigma_1} 
 \begin{pmatrix}
 0 \\
 1 
 \end{pmatrix},
\end{align}
where $\sigma_Z\equiv 2 \hat{T}_Z$ (note $\hat{Q}H_i =0$ for $r \to \infty$).
It is clear that these configurations have a winding number unity for $U(1)_{\mathrm{PQ}}$, $-\cos 2\beta$ for $U(1)_Z$ 
and unity for the $U(1)_{W^1}$ subgroup ($\sigma_1$ rotation) of the gauge symmetry.
Therefore, to cancel the gradient energy from the windings for $U(1)_Z$ and $U(1)_{W^1}$,
the ansatz for the gauge fields are given as
\begin{equation}
  Z_i = \frac{ 2\cos 2\beta }{g_Z}\frac{ \epsilon_{ij} x_j}{r^2 }(1-z(r)), 
 \label{014525_1Oct20}
\end{equation}
\begin{equation}
 W_i^1 = \frac{- 2 }{g}\frac{ \epsilon_{ij} x_j}{r^2 }(1-w(r)),
 \label{180123_22Sep20}
\end{equation}
and $A_i=0$,
where we have used the definitions of the gauge fields Eqs.~\eqref{012344_6Oct20} and \eqref{012356_6Oct20}.
The profile functions for the gauge fields
$w(r)$ and $z(r)$ satisfy 
\begin{equation}
w(0)=z(0)=1,\h{1em} w(\infty)=z(\infty)=0.
\end{equation}

Interestingly, the type-C string has both of the $Z$ and $W$-fluxes for $\tan \beta \neq 1$.
It follows from the ansatz \eqref{014525_1Oct20} and \eqref{180123_22Sep20} that
\begin{equation}
 \Phi_Z = \oint_{r= \infty} dx_i~ Z_i= \frac{-4 \pi \cos 2\beta}{g_Z},
\end{equation}
\begin{equation}
 \Phi_{W^1} = \oint_{r= \infty} dx_i~ W_i^1= \frac{ 4 \pi }{g}\label{015221_1Oct20},
\end{equation}
where the latter flux is independent of $\tan \beta$.

The most important difference from the type-A and type-B strings is that 
the $U(1)_{\mathrm{EM}}$ symmetry is broken inside the type-C string.
This can be seen by using the concrete expression, Eq.~\eqref{235219_22Sep20}, as
$\hat{Q}H_i \neq 0$ ($i=1,2$).
Actually, that is generally inevitable for a configuration with a non-vanishing winding number for charged components,
i.e., a configuration whose asymptotic form is $\exp[i \hat{T}(\theta)] H(\theta=0)$ 
with a non-Abelian generator $\hat{T}(\theta)$ satisfying $[\hat{T}(\theta), \hat{Q}] \neq 0$ \cite{Alford:1990mk,Alford:1990ur}.
($\hat{T}=\sigma_1 \theta$ in our case.)
Due to the winding,
the charged components cannot remain zero,
and they acquire non-zero values inside the string,
leading to the breaking of $U(1)_{\mathrm{EM}}$.
We should note that $U(1)_{\mathrm{EM}}$ is restored at large distances from the string, $r \to \infty$.

\begin{figure}[tbp]
 \centering
\includegraphics[width=0.55\textwidth]{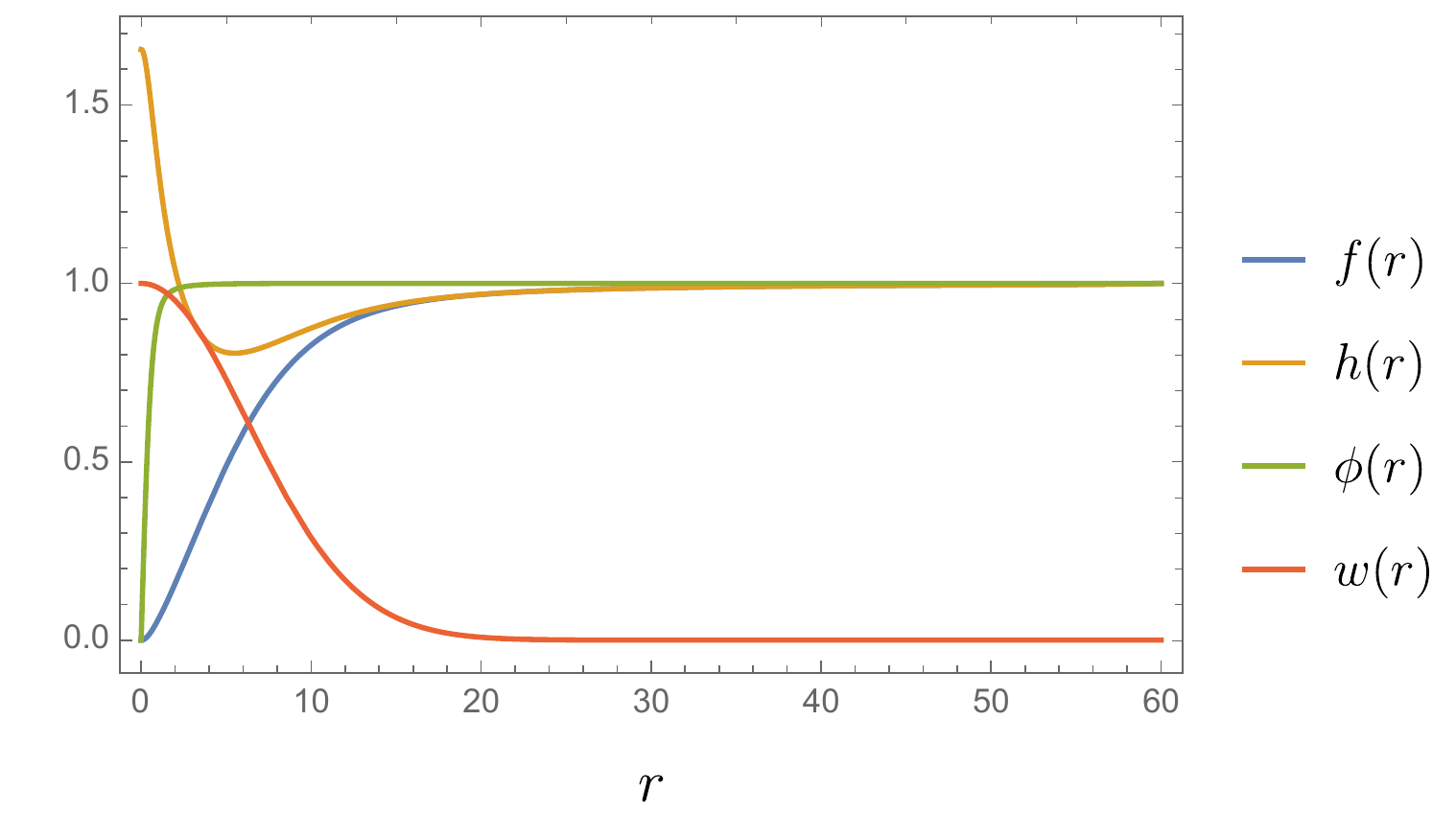} \hspace{0em}
\includegraphics[width=0.43\textwidth]{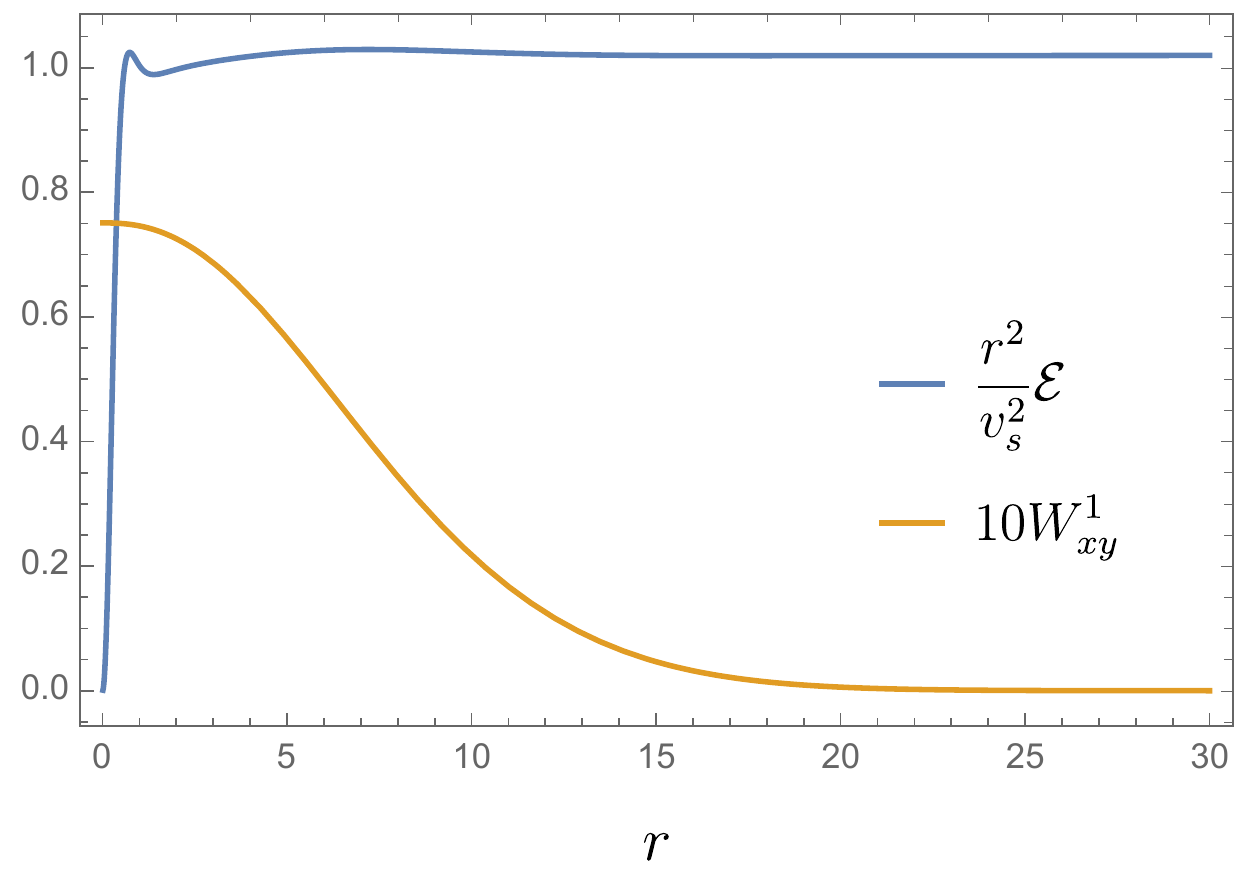}
\caption{
Numerical solution for the type-C string.
We take the same benchmark parameters as Fig.~\ref{044741_30Sep20}.
Also we adopt a length unit as $v_s^{-1}=0.5$.
(left): Plots of profile functions. 
The behavior of the scalar profile functions are the same as one in Fig.~\ref{045818_30Sep20}.
The profile function of the $W^1$ field, $w(r)$, approaches to zero as $r \to \infty$ starting from unity at $r=0$.
(right): Plots of energy density \eqref{193859_3Oct20} divided by $v_s^2/r^2$ 
and the $W^1$-flux density multiplied by $10$.
The tension integrated over $0 \leq r \leq 120 \, v_s^{-1}$ is $140.604$.
The total value of the $W^1$-flux is calculated to $19.3208$, which is consistent with Eq.~\eqref{015221_1Oct20}.
}
\label{013813_1Oct20}
\end{figure}

Let us solve the EOMs with respect to the profile functions 
and calculate the string tension for the type-C string.
Again, we take $m_2^2=\alpha_4=0$ and $\kappa_{1S}=\kappa_{2S}$,
leading to $\tan \beta=1$.
The VEVs are denoted as $v_1=v_2 \equiv v$.
The Higgs doublets have winding numbers for the $U(1)_{W^1}$ gauge subgroup but not for $U(1)_Z$.
Consequently, only the $W^1$-flux is non-zero and confined inside the string.
Substituting the ansatz, the energy density is given by
\begin{align}
 \mathcal{E} =& \frac{v^2}{r^2} \left[r^2 \left(f'(r)^2+h'(r)^2\right)+f(r)^2 (w(r)+1)^2+h(r)^2
   (w(r)-1)^2\right] \n\label{193859_3Oct20} \\
&-m_1^2v^2 (f(r)^2 + h(r)^2) +  
v^4\left[2 (\alpha_2+\alpha_3) f(r)^2 h(r)^2 + (\alpha_1+\alpha_2) (f(r)^4 + h(r)^4)\right] \n \\
&+v^2 v_s^2 \left[2 \kappa  f(r) h(r) \phi(r)^2
 +\kappa_{1S} (f(r)^2+h(r)^2) \phi (r)^2 \right] \n \\
&+\frac{2 w'(r)^2}{g^2 r^2}
+v_s^2 \left(-m_S^2 \phi(r)^2+\lambda_{S} v_s^2 \phi (r)^4\right)
+\frac{v_s^2}{r^2} \left(r^2 \phi '(r)^2+\phi (r)^2\right),
\end{align}
and, the EOMs are given as follows:
\begin{align}
 & f''(r) + \frac{f'(r)}{r} - \frac{(1 + w(r))^2}{r^2} f(r)
 \nonumber\\
 &- \Bigl( 2 (\alpha_1 + \alpha_2)v^2  f(r)^2+ 2 (\alpha_2 + \alpha_3) v^2 h(r)^2 + \kappa_{1S} v_s^2 \phi(r)^2 -m_1^2 \Bigr) f(r) - \kappa v_s^2 h(r) \phi(r)^2 =0,
 \\
 &h''(r) + \frac{h'(r)}{r} - \frac{(-1 + w(r))^2}{r^2} h(r)
 \nonumber\\
 &- \Bigl( 2 (\alpha_1 + \alpha_2) v^2 h(r)^2 + 2 (\alpha_2 + \alpha_3) v^2 f(r)^2 + \kappa_{1S} v_s^2 \phi(r)^2 -m_1^2 \Bigr) h(r) - \kappa v_s^2 f(r) \phi(r)^2 =0,
 \\
 &\phi''(r) + \frac{\phi'(r)}{r} - \frac{\phi(r)}{r^2}
 \nonumber\\
 & - \Bigl( 2 \lambda_S v_s^2 \phi(r)^2 + \kappa_{1S} v^2 (f(r)^2 + h(r)^2) + 2 \kappa v^2 f(r) h(r) \Bigr) \phi(r)=0,
\end{align}
\begin{align}
 &w''(r) - \frac{w'(r)}{r} - \frac{g^2 v^2}{2} f(r)^2 (1 + w(r)) - \frac{g^2 v^2}{2} h(r)^2 (-1 + w(r) )=0.
\end{align}

The obtained numerical solutions are shown in Fig.~\ref{013813_1Oct20}.
We take the same parameter choice as Eq.~\eqref{052110_30Sep20}
and set the VEV for $S$ as $v_s=10 \, v$.
The shapes of the profile functions, the energy density and the flux density are almost similar to those of the type-B string (Fig.~\ref{045818_30Sep20}).
This can be understood by the argument on non-Abelian moduli in 2HDM in Ref.~\cite{Eto:2018tnk}.
That is, when $\tan \beta =1$, the two ansatz for the Higgs fields (Eqs.~\eqref{194425_3Oct20} and \eqref{235219_22Sep20}) are related by the $SU(2)_C$ custodial transformation:
\begin{equation}
 H \to U^\dagger H U
\end{equation}
with $U=\exp \left[i \frac{\pi}{4}\sigma_2\right]$.
This symmetry is respected in the potential $V(H,S)$ when $m_2^2=\alpha_3=\alpha_4=0$ and $\kappa_{1S}=\kappa_{2S}$, 
but is explicitly broken in the gauge sector because of $g'\neq 0$.
Thus the shapes and hence the string tension are slightly different between them.

While the above ansatz has the $Z$- and $W^1$-fluxes in the string,
there is also a string in which the $W^2$-flux (and also mixtures of them in general) is confined.
Due to the $U(1)_{\mathrm{EM}}$ symmetry in the Lagrangian, they have the degenerated tension.
Since the property is almost the same as the one we studied above,
we do not consider them in this paper.

\subsection{String tensions}

The vortex strings we have considered above have the same winding number (unity) associated with the $U(1)_{\mathrm{PQ}}$ symmetry.
Because all possible configurations in the theory are classified into topological sectors characterized by the non-trivial first homotopy group $\pi_1(U(1)_{\mathrm{PQ}})=\mathbb{Z}$, 
the above fact means that they are in the same topological sector with the topological charge $1 \in \pi_1(U(1)_{\mathrm{PQ}})=\mathbb{Z}$
and that they can continuously deform to each others.
Since their string tensions (energy per length unit) are generically not degenerated,
heavier strings decay into the lightest one, 
which does not decay any further and is a stable solution to the EOMs.%
We here study the string tensions, i.e., stability of the strings.

For simplicity, we focus on a case with $\tan \beta=1$.
This is realized when 
$m_2^2=\alpha_4 = 0$ in the Higgs potential \eqref{213928_10Sep20}
and $\kappa_{1S}=\kappa_{2S}$ in the mixing term \eqref{183300_19Sep20}.
In this case, as stated above, the type-A string does not have the $Z$-flux 
and the type-B one has the winding number unity for $U(1)_Z$.
In addition, the type-C string does not have the $U(1)_Z$ winding but does for $U(1)_{W^1}$.
It may seem that the type-A one is lighter than type-B and type-C ones since the latter two have the $Z$ and $W$-fluxes.
However, this is not the case when the potential energy is more dominant than that of the gauge sector.
Indeed, both the profile functions for the doublets in the type-A string vanish on the core,
and thus that leads to a larger amount of the potential energy than those of the type-B and type-C.

On the other hand, the difference of the tensions between the type-B and type-C strings is controlled by the parameter $\alpha_3$.
As explained in the last subsection, in the case $\tan \beta=1$, the type-C string has only the $W^1$-flux,
and the two strings are related by the custodial $SU(2)_C$ transformation.
If $\alpha_3=0$ (and $m_2^2=\alpha_4=0$), 
the Higgs potential $V_H$ respects this symmetry 
but the gauge sector does not due to the $U(1)_Y$ coupling constant $g' \neq 0$.
This slightly lifts up the tension of the type-C because of $g_Z =\sqrt{g^2 + g'{}^2}> g$.
The non-zero value of $\alpha_3$ can change this relation.
In Refs.~\cite{Eto:2018tnk,Eto:2020hjb}, 
it is shown that, in 2HDM, the smaller (larger) value of $\alpha_3$ tends to make
the tension of the string with the $W$-flux heavier (lighter) than that of the string with the $Z$-flux.
Thus, in our case, we expect that the type-C string is lighter than the type-A and type-B ones when $\alpha_3$ is larger than a critical value depending on other parameters.

Keeping $\kappa_{1S}=\kappa_{2S}$, we take the benchmark parameters as
\begin{equation}
\alpha_1=1, \h{1em}\lambda_S=1, \h{1em}\kappa=-2\left(\frac{v}{v_s}\right)^2, \h{1em} v_s = 10 \, v, \h{1em} m_2^2=\alpha_4 = 0, 
\label{045724_2Oct20}
\end{equation}
and take $\alpha_2$ such that the lightest scalar mass $m_{h_1}^2$ is equal to the SM Higgs mass $(125\, \mathrm{GeV})^2$.
We use a length unit $v_s^{-1}=0.5 $.
The remaining two parameters $\alpha_3$ and $\kappa_{1S}$ are taken as free parameters.
The tensions $T$ are calculated over $0 \leq r \leq L$ with the IR cutoff $L=120\,v_s^{-1} $.
Note that, although each tension depends on the IR cutoff as $\sim \log L$,
the differences do not, so that we can compare them as far as $L$ is fixed.

\begin{figure}[tbp]
 \centering
\includegraphics[width=0.47\textwidth]{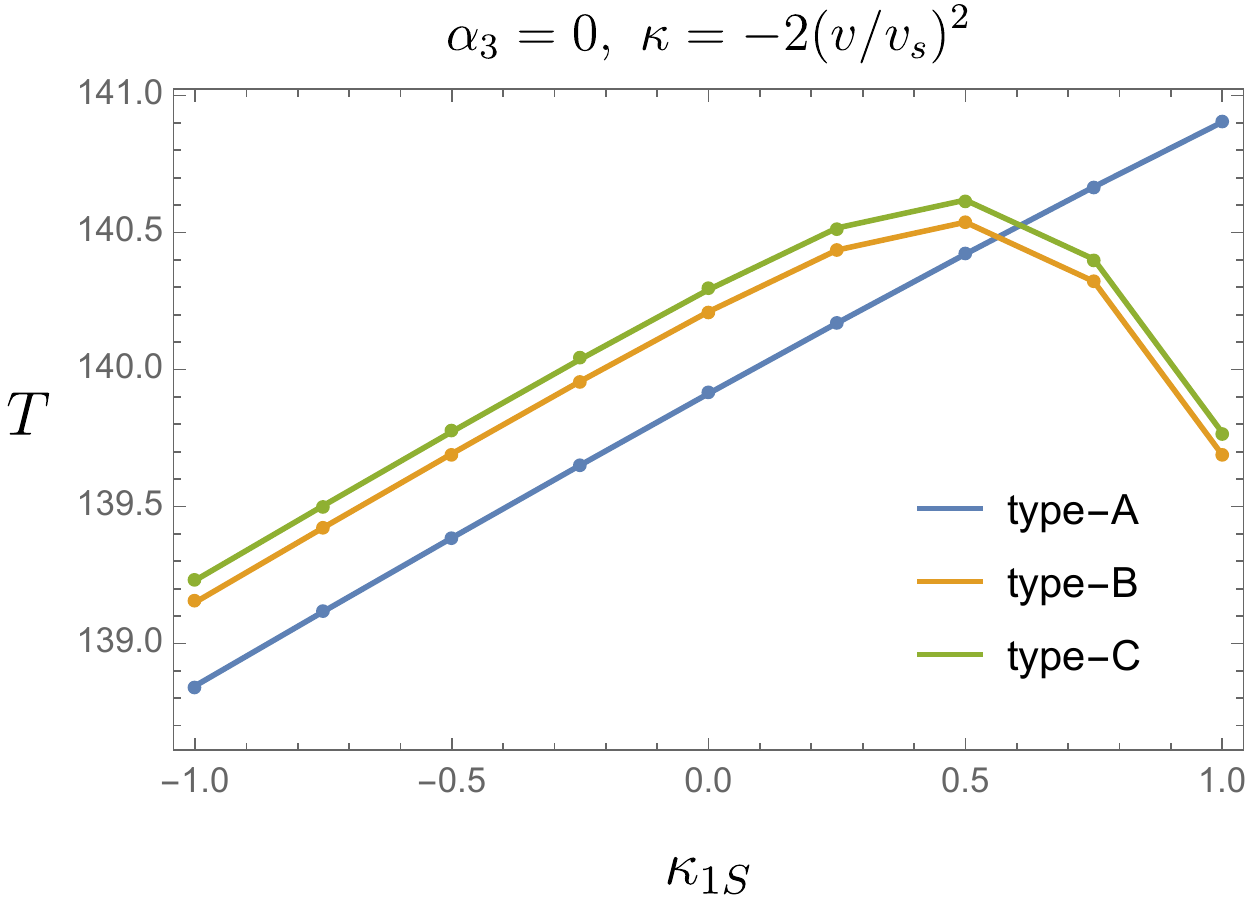} \hspace{0.5em}
\includegraphics[width=0.47\textwidth]{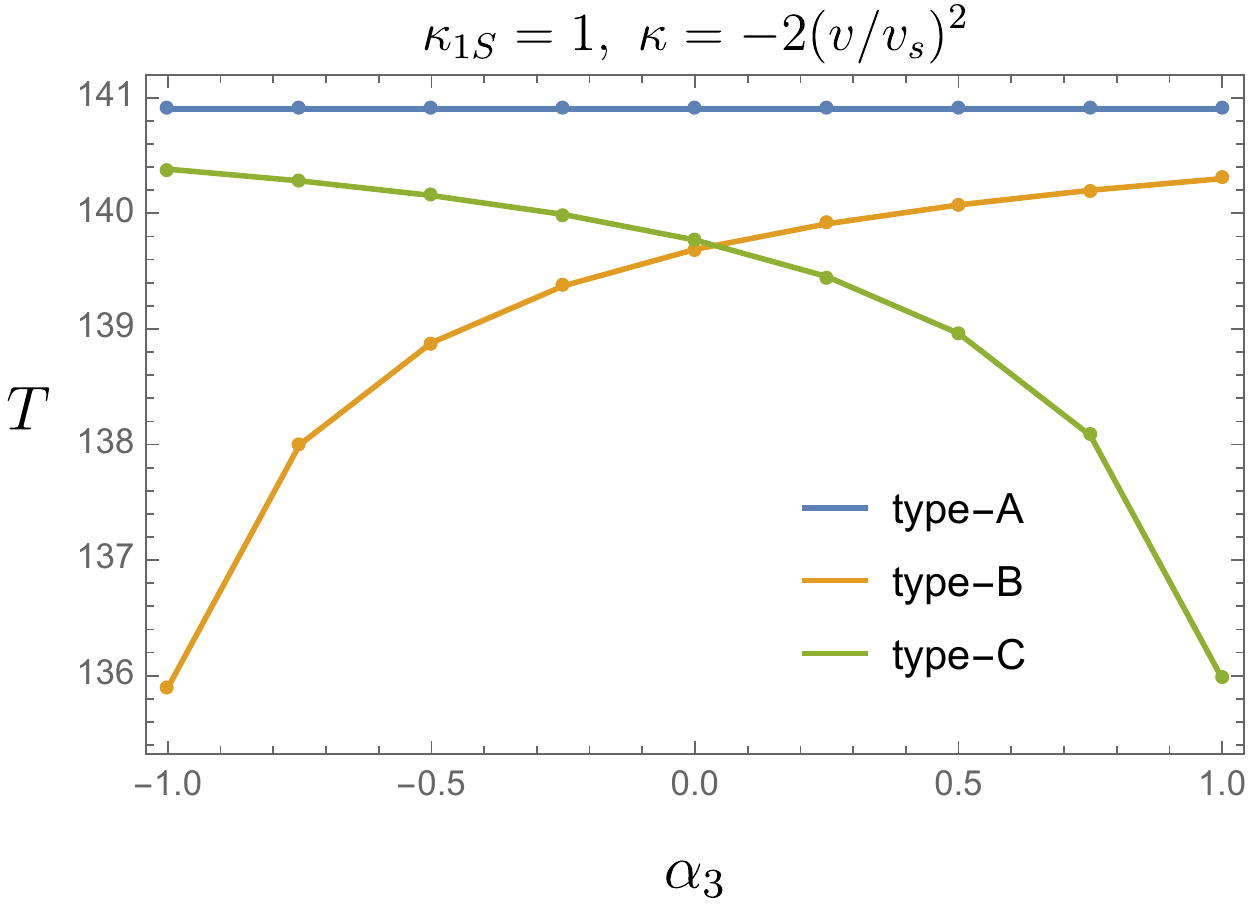}
\caption{
String tensions of the type-A, type-B and type-C strings.
The parameters are taken as Eq.~\eqref{045724_2Oct20}
and the tension is calculated by integration over $0\leq r \leq 120 ~v_s^{-1}$.
(left): $\alpha_3$ is fixed as 0 and $\kappa_{1S}$ is taken as a free parameter.
The tension of the type-A string increases as $\kappa_{1S}$ becomes larger,
but those of the type-B and type-C ones change to decrease for $\kappa_{1S}\gtrsim 0.5 $.
(right): $\kappa_{1S}$ is fixed as $1.0$ and $\alpha_3$ is taken as a free parameter.
As is expected, the tension of the type-B (type-C) strings decreases (increases) as $\alpha_3$ becomes larger.
The type-C string becomes the lightest one for $\alpha_3 \gtrsim 0$.
}
\label{041121_2Oct20}
\end{figure}

Fig.~\ref{041121_2Oct20} shows the relation of the string tensions between the three strings.
In the left panel, we fix $\alpha_3=0$ and scan $\kappa_{1S}$ in the range $-1 \leq \kappa_{1S} \leq 1$.
The tension of the type-A string increases as $\kappa_{1S}$ becomes larger,
but those of the type-B and type-C ones change to decrease for $\kappa_{1S}\gtrsim 0.5 $.
It can be seen that the difference of the tensions between type-B and type-C is independent of $\kappa_{1S}$, 
which is reasonable because it is controlled only by the $SU(2)_C$ breaking parameters $g'$ and $\alpha_3$ as stated above.

In the right panel, we fix $\kappa_{1S}=1.0$ and scan $\alpha_3$ in the range $-1 \leq \alpha_3 \leq 1$.
As is expected, the tension of the type-C (type-B) strings increases (decreases)  as $\alpha_3$ increases.
In this parameter choice, that of the type-A string is almost constant, but this tendency depends on the parameters in general.
The type-C string becomes the lightest one for $\alpha_3 \gtrsim 0$.

Before closing this section, 
we stress that the type-C string can be the lightest and stable string for rather wide parameter space.
Although we have concentrated on the case of $\tan \beta=1$, it would not be crucial.
Therefore, the axion string produced by the breaking of $U(1)_{\mathrm{PQ}}$ in the early universe 
necessarily becomes the type-C electroweak axion string after the electroweak phase transition, depending on the parameters in the DFSZ model.
In the string, the $U(1)_{\mathrm{EM}}$ symmetry is spontaneously broken.
This property causes an interesting phenomenon on the string, superconductivity of vortex strings,
as we see in the next section.

\section{Superconducting DFSZ string}
\label{sec:superconducting}

It is known \cite{Witten:1984eb} that cosmic strings can be superconductors, 
i.e., the electric current can flow along a string without resistance,
when the electromagnetic gauge symmetry is broken inside the string core.
Superconducting strings are often realized 
by using scalar fields that develop non-zero VEVs only inside the strings
or fermionic fields whose gapless modes are confined on the string.
Further, non-Abelian vortex strings, in which charged 
particles such as charged vector bosons are condensed, 
can also support superconductivity~\cite{Alford:1990ur,Alford:1990mk}.
In this section, we show that the type-C string discussed in
Sec.~\ref{sec:electroweakaxionstrings} can be a superconducting string. 

In the type-C string, the $W^1$-flux is confined,  
and the $U(1)_\mathrm{EM}$ symmetry is broken by the charged Higgs components and the $W^1$ gauge field.
Corresponding to the breaking of $U(1)_\mathrm{EM}$, the string has a $U(1)_\mathrm{EM}$ moduli parameter,
which is a flat direction around the string configuration.
The existence of the moduli ensures that a $(z,t)$-dependent fluctuation in the direction of the moduli
is a zero mode (massless excitation) in the string background and can travel on the string with the speed of light.
It can carry an electric current without resistance, resulting in a supercurrent.
We can rephrase this explanation into a more concrete one.
The breaking of $U(1)_\mathrm{EM}$ means that the charged components of the Higgs field gets the non-zero VEV inside the string, $\hat{Q}H \neq0$,
and that they can play a similar role to the charged scalar field in Ref.~\cite{Witten:1984eb}.

In addition, the string can carry large electric current
which induces a large magnetic interaction between the strings.
That may affect the cosmological evolution of the string in the
DFSZ axion model.
In the following analysis, we assume $\tan \beta=1$ for simplicity, 
but it is irrelevant to the argument on superconductivity.

\subsection{Zero modes along the string}
Let $\ol{S}$, $\ol{H}$, $\ol{W}_\mu$ and $\ol{Y}_\mu$ 
be the background configuration for the type-C string given by Eqs.~\eqref{235219_22Sep20} and \eqref{180123_22Sep20}. 
Note that $\ol{Z}_\mu=0$ due to $\tan \beta=1$.
To find a zero mode excitation,
we consider the $(z,t)$-modulated ``gauge transformation'' around the type-C string:
\begin{align}
 S & = \ol{S}, \\
 H & = \exp\left[ i\eta(z,t) \chi(r,\theta)\right]~\ol{H}~\exp\left[i\eta(z,t) \xi(r,\theta) \f{\sigma_3}{2}\right], \\
 W_\mu & = \exp[ i\eta(z,t) \chi(r,\theta)] \left( \ol{W}_\mu - \frac{i}{g} \delta_\mu ^j \partial_j\right) \exp[ -i\eta(z,t) \chi(r,\theta)],\\ 
 Y_\mu & = \ol{Y}_\mu + \frac{1}{g'} \delta_\mu ^j\,  \eta(z,t) \partial_j\xi(r,\theta)
\end{align}
where $\chi=\chi^a \sigma_a/2$ and $j = r,\theta$.  
This is not a mere gauge transformation unless $\eta$ is independent of $z$ and $t$,
but is a $(z,t)$-dependent physical excitation described by $\eta(z,t)$, $\xi(r,\theta)$ and $\chi^a(r,\theta)$.

Instead of the above expressions, for later use, we analyze an alternative ansatz that is obtained by performing the $SU(2)_W \times U(1)_Y$ gauge transformation with the gauge parameters $(\eta \chi, \eta \xi)$.
The transformed ansatz is given by
\begin{align}
 S&= \ol{S}, \label{151514_24Sep20} \\
 H&= \ol{H}, \\
 W_\mu &= \ol{W}_\mu + \delta W_\mu, \\
 Y_\mu & =\ol{Y}_\mu + \delta Y_\mu\label{151518_24Sep20}
\end{align}
with
\begin{align}
 \delta W_\mu 
 &= \frac{1}{g}\delta_\mu^\alpha \,\chi \partial_\alpha \eta, \\
 \delta Y_\mu 
 &= - \frac{1}{g'}\delta_\mu^\alpha \,\xi \partial_\alpha \eta ,
\end{align}
where $\alpha=t,z$.
For these ansatz \eqref{151514_24Sep20}-\eqref{151518_24Sep20}, 
the field strength tensors are given by
\begin{align}
 W_{\mu\nu} &= \ol{W}_{\mu\nu} + \ol{D}_\mu \delta W_\nu - \ol{D}_\nu \delta W_\mu,\\
 Y_{\mu\nu} &= \ol{Y}_{\mu\nu} + \partial_\mu \delta Y_\nu - \partial_\nu \delta Y_\mu,
\end{align}
where $\ol{W}_{\mu\nu}$, $\ol{Y}_{\mu \nu}$ and $\ol{D}_\mu$ are 
the field strengths and the covariant derivative consisting of 
the background gauge configurations $\ol{W}_\mu$ and $\ol{Y}_\mu$.

The linearized EOMs for the excitation $\eta, \chi^a $ and $\xi$
are obtained by substituting the ansatz into the full EOMs,
\begin{align}
\left( D_\nu W ^{\nu\mu} \right)^a &= - j^{\mu,a}_{W},\label{160020_24Sep20} \\
\partial_\nu Y ^{\nu\mu} &=  - j^\mu_{Y},\label{165528_24Sep20} \\
D_\mu D ^\mu H &= -\frac{\delta V(H,S)}{\delta H^\dagger},\label{172912_24Sep20}
\end{align}
where $j_W^{\mu,a}$ and $ j_Y^{\mu}$ are the $SU(2)_W$ and $U(1)_Y$ currents:
\begin{align}
 j_W^{\mu,a} &= \f{i}{2}g ~\mathrm{Tr} \left[H ^\dagger \sigma^a D ^\mu H - (D^\mu H) ^\dagger \sigma^a H\right],\label{051758_25Sep20}\\
 j_Y^{\mu} &= - \f{i}{2}g' ~\mathrm{Tr} \left[\sigma_3 H ^\dagger D ^\mu H - (D^\mu H) ^\dagger H \sigma_3 \right].\label{051805_25Sep20}
\end{align}
Then we obtain the following equations 
from Eqs.~\eqref{160020_24Sep20} and \eqref{165528_24Sep20}
(see Appendix \ref{041710_25Sep20} for the derivation),
\begin{align}
 \partial^\alpha \eta \left(\ol{D}_j  \ol{D}^j \chi \right)^a 
&= \f{-g^2}{2} \partial^\alpha \eta  \left(\chi^a  \mathrm{Tr}|\ol{H}|^2
 + \xi \mathrm{Tr} \left[\ol{H}^\dagger \sigma_a \ol{H} \sigma_3  \right] \right), \label{163119_24Sep20}\\
  \partial^\alpha \eta \partial_j \partial^j \xi
&= \f{-g'^2}{2} \partial^\alpha \eta  \left(\xi  \mathrm{Tr}|\ol{H}|^2
 + 2 \mathrm{Tr} \left[\ol{H}^\dagger \chi \ol{H} \sigma_3  \right] \right),\label{035453_25Sep20}\\
 \ol{D}^j \chi  ~\partial^\alpha \partial_\alpha \eta & =0, \label{164640_24Sep20}\\
 \partial^j \xi  ~\partial^\alpha \partial_\alpha \eta &=0, \label{174012_24Sep20}
\end{align}
and from the EOM for $H$~\eqref{172912_24Sep20}
\begin{equation}
 \partial^\alpha \partial_\alpha \eta \left(2 \chi \ol{H} + \xi \ol{H} \sigma^3 \right) = 0.\label{174016_24Sep20}
\end{equation}
We have used the fact that
the background configurations $\ol{H}$, $\ol{W}_\mu$ and $\ol{Y}_\mu$
solve the EOMs. 

The above equations \eqref{164640_24Sep20}, \eqref{174012_24Sep20} and \eqref{174016_24Sep20} are satisfied with
\begin{equation}
  \partial^\alpha \partial_\alpha  \eta =
  (\partial_t^2 -\partial_z^2) \eta =0 ,
\end{equation}
which is a (1+1)-dimensional wave equation.
This has the zero mode solutions $\eta=\eta^+(z+t)$ and
$\eta=\eta^-(z-t)$ with some functions $\eta^\pm$, and the general
solution can be written by a linear combination of these modes. When
one is particularly 
interested in the static case, the $t$-independent solution is given
by 
\begin{align}
  \eta(z) = \omega z, \h{2em} (\omega : \text{const.})
\end{align}
implying 
the constant current along the $z$ direction.
On the other hand, the radial and angular dependence of the excitations is determined by Eqs.~\eqref{163119_24Sep20} and \eqref{035453_25Sep20}.
We can see that there are four independent zero modes corresponding to
the solutions $\chi^a$ ($a=1,2,3$) and $\xi$.
This is understood from the fact that the whole symmetry of
$SU(2)_W \times U(1)_Y$ (even 
$U(1)_\mathrm{EM}$) is broken inside the string.
In other words, four zero modes can induce the $SU(2)_W$ and $U(1)_Y$
currents on the string.

However, only one of them induces the $U(1)_\mathrm{EM}$ current, 
which generates two-dimensional Coulomb (magnetic) potential far from
the string.
To illustrate this, let us consider the asymptotic behavior of $\chi$ and $\xi$ at $r \to \infty$.
At infinity, the $U(1)_\mathrm{EM}$ is restored,
in which the background configuration $\ol{H}$ satisfies \cite{Eto:2020opf}
\begin{equation}
 \ol{H} \sigma ^3 + n^a \sigma^a \ol{H} =0,\h{2em}
 n^a = - \frac{\mathrm{Tr}\left(\sigma^3 \ol{H}^\dagger \sigma^a \ol{H}\right)}
{\mathrm{Tr}|\ol{H}|^2},
\end{equation}
and $(\ol{D}_\mu n)^a =0$ and $\mathrm{Tr}|\ol{H}|^2=2 \, v^2$.
Using these conditions,
the Poisson-like equations \eqref{163119_24Sep20} and \eqref{035453_25Sep20} become
\begin{align}
 \left(\ol{D}_j  \ol{D}^j \chi \right)^a 
&= -g^2 v^2  \left(\chi^a - n^a \xi \right), \\
 \partial_j \partial^j \xi
&= -g'^2 v^2 \left(\xi - \chi^a n^a \right) , 
\end{align}
which describe the long-range behavior on the $xy$ plane
only when $\chi^a - \xi n^a =0$.
Thus, we have the two-dimensional Laplace equation:
\begin{equation}
 \frac{1}{r}\partial_r (r \partial_r \xi) =0\label{072138_25Sep20}
\end{equation}
with $\chi^a = \xi n^a$ for $\xi$ being rotationally 
invariant ($\partial_\theta \xi=0$).
The asymptotic solution of Eq.~\eqref{072138_25Sep20} behaves as $\xi \sim \log r $.
Substituting this into the expressions of gauge fields gives
\begin{equation}
 \delta W_z^a \sim \frac{\omega}{g} n^a \log r,
\h{2em} \delta Y_z \sim - \frac{\omega}{g'} \log r,
\end{equation}
where we have taken the normalization of $\xi$ such that $\xi \to \log r$ for $r \to \infty$.
We find the form of the $U(1)_\mathrm{EM}$ field strength 
\begin{align}
 F_{rz}^\mathrm{EM} &= - \sin \theta_W n^a W_{rz}^a + \cos \theta_W Y_{rz}, \\
 &= - \sin \theta_W n^a \partial_r \delta W_{z}^a + \cos \theta_W \partial_r \delta Y_{z}, \\
 & \sim - \f{\omega}{er} \hspace{1.5em} (r \to \infty).\label{200133_5Oct20}
\end{align}
This is nothing but the magnetic long-range force on the two-dimensional $xy$ plane.
Correspondingly, the total amount of $U(1)_\mathrm{EM}$ current $J_\mathrm{EM}$ along the string is estimated from Eq.~\eqref{200133_5Oct20} as
\begin{equation}
 J_\mathrm{EM} \equiv - 2 \pi r F_{rz}^\mathrm{EM} \sim \f{2\pi \omega}{e}.\label{001103_1Feb21}
\end{equation}

\subsection{Current quenching and string interaction}
In the above argument, 
it may seem that the magnitude of the current is given by the
parameter $\omega$ and  can be taken arbitrarily large.
However this is not the case
since we have ignored the backreaction from the zero modes to the
background fields $\ol{S}$, $\ol{H}$, $\ol{W}_\mu$ and $\ol{Y}_\mu$
by linearizing the EOMs. To examine the backreaction, 
we look at the following term in the Lagrangian:
\begin{align}
  \mathcal{L} &\supset -\mathrm{Tr} |-ig \delta W_z \ol{H} + i \tfrac{g'}{2} \ol{H}\sigma_3\delta Y_z |^2 \\ 
&= - \omega^2 \mathrm{Tr}|\chi  \ol{H} + \xi \ol{H}\tfrac{\sigma_3}{2} |^2 \label{101825_30Sep20} ,
\end{align}
which is obtained by substituting the string ansatz and the solution
for $\eta(z)$.
This term induces a positive squared mass for the $SU(2)_W \times
U(1)_Y$ charged components of $\ol{H}$.
For $\omega\to\infty$,  
the charged components vanish, $\chi \ol{H} + \xi \ol{H}\tfrac{\sigma_3}{2} \to 0$, 
which decreases the $SU(2)_W$ and $U(1)_Y$ currents from the Higgs field, $j_W ^{z,a}$ and $j_Y^z$ (\eqref{051758_25Sep20} and \eqref{051805_25Sep20}).
Correspondingly, the right hand sides of Eqs.~\eqref{163119_24Sep20} and \eqref{035453_25Sep20} vanish everywhere,
and they have only a trivial solution $\xi=\chi^a =0$.%
\footnote{A solution behaving like $\sim \log r$ for all $r$ is singular at $r=0$.}
Thus, the backreaction from an extremely large $\omega$ reduces the amount of the current.
Such behavior is known as the current quenching \cite{Witten:1984eb}.

Now, let us estimate the maximum value of the current.
For the $U(1)_\mathrm{EM}$ zero mode, $\chi^a = \xi n^a$,
the mass term \eqref{101825_30Sep20} reads
\begin{equation}
 -(\omega\xi)^2 \mathrm{Tr}|\hat{Q}\ol{H}|^2 = -(\omega\xi)^2 \frac{v^2}{2}(f-h)^2,\label{023518_1Oct20}
\end{equation}
where $\hat{Q}$ is the $U(1)_\mathrm{EM}$ generator defined in Eq.~\eqref{010004_1Oct20},
and we have used the concrete expression of $\ol{H}$ for the type-C string, Eq.~\eqref{235219_22Sep20}.
On the other hand, using $\kappa_{1S}=\kappa_{2S}$ for $v_1=v_2=v$, the mass terms for $f$ and $h$ that are originally present in the Lagrangian are
\begin{equation}
-\mathcal{L} \supset \frac{v^2}{r^2}\left[(1+w)^2 f^2 +(1-w)^2 h^2 \right] + (- m_1^2 +\kappa_{1S}v_s^2  \phi^2 )(f^2+h^2) .
\end{equation}
Due to the backreaction term Eq.~\eqref{023518_1Oct20}, 
the two-by-two mass matrix $M^2$ for $(f,h)$ is changed and not diagonal.
It is sufficient to consider the signs of the eigenvalues of $M ^2$.
If they are positive inside the string, then $f$ and $h$ tend to vanish, and hence $(f-h)^2 \propto \mathrm{Tr}|\hat{Q}H|^2=0$, 
which means that the $U(1)_\mathrm{EM}$ symmetry is restored even inside the string.
On the other hand, if one of them is negative, the quenching is not significant
and the $U(1)_\mathrm{EM}$ symmetry is still broken.
Then the current can be increased with $|\omega|$.

Inside the string, $r \lesssim v^{-1}$, 
the determinant of the matrix $M^2$ is calculated as
\begin{equation}
 \mathrm{det}M ^2 \sim m_1^4 + \frac{4}{r^2}\left( \omega ^2 \xi^2 - m_1^2  +\kappa_{1S} \phi^2\right) + \left(-  m_1^2  +\kappa_{1S} v_s^2 \phi^2 \right)\omega^2 \xi^2 ,\label{224116_6Oct20}
\end{equation}
where we have used $w(r) \sim 1$ there.
Note that
the mass matrix and hence the determinant vary with the radius $r$ in and out the string core as
\begin{equation}
\mathrm{det}M^2 \sim \begin{cases}
  v_s^4 + \frac{4}{r^2}\left( \omega ^2 \xi^2 - v^2 \right) -v^2 \omega^2\xi^2 
 & \text{for}~r\gtrsim v_s^{-1},\\
  v_s^4 + \frac{4}{r^2}\left( \omega ^2 \xi^2 - v_s^2 \right) -v_s^2 \omega^2\xi^2 
 & \text{for}~ r \lesssim v_s^{-1}.
\end{cases}
\end{equation}
Clearly, there is a critical value of $|\omega \xi|$,
for which the sign of the term proportional to $r^{-2}$ changes from negative to positive.
For a value smaller than the critical one, there exists a region where the determinant is negative,
avoiding the current quenching.
The critical value of $|\omega\xi|$ is $\sim v_s$ around the core.
Then the current magnitude becomes maximum for this value.
Note that $\xi$ is of order unity inside the core to be connected with the asymptotic form $\xi \to \log r$.
Thus, $|\omega| \sim v_s$ in this case.
Note that the definition of the $U(1)_\mathrm{EM}$ moduli may be different a bit from $\chi^a = \xi n^a$
inside the string \cite{Eto:2020opf}.
The difference is however subleading and negligible for the maximum amount of the supercurrent.

Finally, we consider the tension of the string 
(energy per unit length). When $\omega=0$, the string reduces to the type-C string 
and the tension $T$ is dominated by the gradient energy from the axion:
\begin{equation}
 T(\omega=0) \sim 2\pi \int^L r dr |\partial _i S|^2 \sim 2 \pi v_s^2 \log L ,
\end{equation}
with $L$ being an IR cutoff.
The logarithmic divergence is natural because $U(1)_\mathrm{PQ}$ is a global symmetry.
On the other hand, for $\omega \neq 0$,
there should be an additional contribution to the tension from the magnetic field induced by the $U(1)_\mathrm{EM}$ current:
\begin{align}
 T(\omega) &= T(0) + T _\mathrm{EM} 
\end{align}
where
\begin{equation}
 T _\mathrm{EM} \sim 2 \pi \int ^L r dr \left(F_{rz}^\mathrm{EM}\right)^2 \sim \f{2 \pi \omega^2}{e^2} \log L .
\end{equation}
Thus the current also induces a logarithmically divergent energy,
which is comparable to that from the gradient term, for the maximal
current $|\omega| \sim v_s$ discussed above.

This result provides an interesting sight for the interaction between the strings.
Let us consider two superconducting strings that are well separated in the $xy$ plane. 
They are assumed to have the same winding for $U(1)_\mathrm{PQ}$ and $U(1)_{W^1}$,
and contain the superconducting currents with the same sign.
As is well-known, the gradient energy of $S$ gives a repulsive force $\sim v_s^2 / R_0$ with $R_0$ being the distance between them.
However, the magnetic interaction induced from the supercurrent provides an attractive force,
\begin{equation}
 F = - \f{(J _\mathrm{EM})^2}{2 \pi R_0} \sim - \f{2\pi\omega^2}{e^2 R_0}.
\end{equation}
Therefore, the superconducting strings can receive the attractive force
overcoming the repulsive one with $|\omega| \sim v_s$.

\subsection{Y-junction formation}
\begin{figure}[t]
\centering
\includegraphics[width=0.7\textwidth]{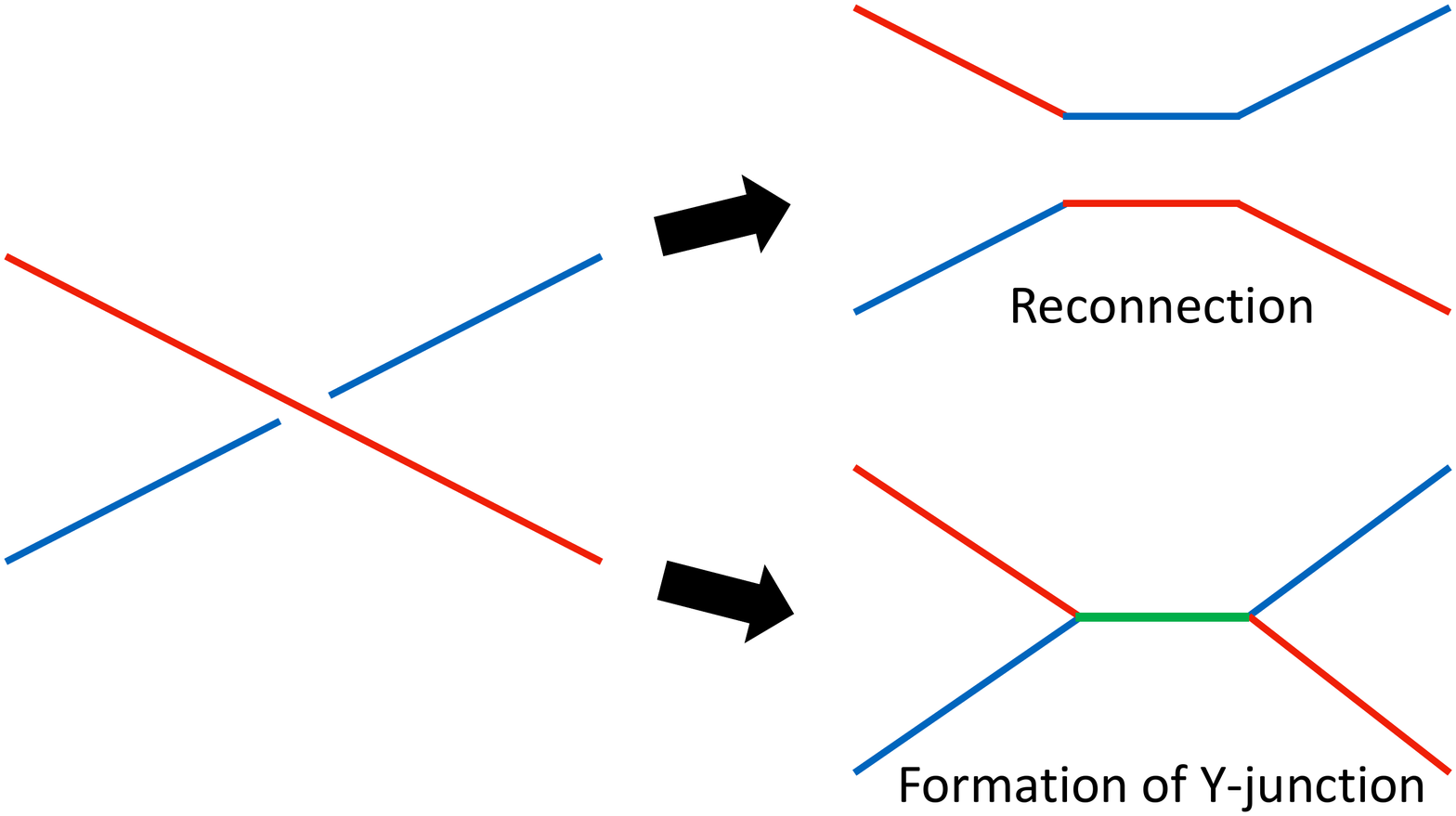}
\caption{
Reconnection process and formation of a Y-junction.
}
\label{142848_24Feb21}
\end{figure}

In usual cases, a pair of axion strings reconnects with a probability of the order of unity when they collide.
Thanks to this reconnection process, a network of strings in the early universe produces small string loops, 
which soon shrink and disappear by emitting radiations,
and sufficiently looses the energy if the network is dense.
The time evolution of the energy density approaches the so-called scaling behavior,
and the total energy density of the universe is not dominated by the string network.
However, the attractive interaction discussed above, which is induced by superconducting current, can change this picture drastically.
In particular, it is known that an attractive interaction between strings could form the Y-junction \cite{Bettencourt:1996qe,Bettencourt:1994kc,Copeland:2006eh,Copeland:2006if,Salmi:2007ah,Bevis:2008hg,Bevis:2009az,Hiramatsu:2013yxa,Hiramatsu:2013tga}, 
which is a bound state of two strings (see Fig.~\ref{142848_24Feb21}).
If Y-junctions are formed frequently, 
that reduces the effective reconnection probability and makes non-trivial whether the string network evolves to the scaling
solution.
As the first step to study the Y-junctions of the axion strings,
we present a rough estimation of the formation probability of Y-junctions in this section.

Firstly, let us assume the presence of a primordial magnetic field (PMF) after the electroweak phase transition.
PMF is well motivated as an origin of intergalactic magnetic fields
in order to explain the galactic magnetic fields observed today via the dinamo mechanism~\cite{Parker:1955zz}.
(For a review, see, e.g., Ref.~\cite{Durrer:2013pga}.)
There are the observational lower and upper bounds on the present strength of intergalactic magnetic fields 
for a coherent length $\lambda \gtrsim 0.1 \, \mr{Mpc}$ as
$3 \times 10^{-16} \, \mathrm{G} \lesssim B_0 \lesssim 10^{-11} \, \mathrm{G}$,
where the lower one is set by the non-observation of secondary photons from the emission of highly collimated gamma rays by blazars~\cite{Neronov_2010}
and the upper one is from the CMB observations~\cite{Jedamzik_2019}.
For simplicity, we further assume that the power spectrum of PMF is scale invariant 
and almost coherent over the entire Hubble horizon,
which could be realized for the inflationary magnetogenesis \cite{Durrer:2013pga}.

We then show that
PMF can induce large superconducting currents on the strings.
Such large currents are sufficient to form the Y-junctions.
The scale invariant PMF evolves as
\begin{equation}
 B(t) \propto a(t)^{-2} \, ,
\end{equation}
where $B(t)$ and $a(t)$ are the magnetic field strength and the scale factor at the cosmological time $t$, respectively.
It is convenient to define a ratio of the energy density of PMF ($\rho_B$) to that of photon ($\rho_\gamma$),
\begin{equation}
 \epsilon \equiv \frac{\rho_B}{\rho_\gamma} \sim B(t)^2 G_N t^2,
\end{equation}
with $G_N$ being the Newton constant.
The ratio $\epsilon$ is a constant as the universe expands.
If we take the upper limit on $B_0$, we have $\epsilon \sim 10^{-11}$.

In the early universe, the superconducting strings (Type-C EW axion strings) move with velocity $v_\mathrm{str}$ ($\sim \mathcal{O}(1)$) in the presence of PMF
and hence feel the electric fields $E \sim B(t) v_\mathrm{str}$,
which induce the superconducting current,
\begin{equation}
 J_\mr{PMF} \sim e^2 B(t) \xi v_\mathrm{str} \, ,
\end{equation}
 with $\xi$ being a string typical length.
Since there are no superconducting strings before the electroweak phase transition,
we can assume that the scaling property $\xi \sim t$ holds at least just after the phase transition,
and obtain 
\begin{equation}
 J_\mr{PMF} \sim 10^{12} \left(\frac{\epsilon}{10^{-11}}\right) ^{1/2}\, \mr{GeV}  \sim 10^{12} \left(\frac{B_0}{10^{-11} \, \mr{G}}\right)\, \mr{GeV} \, .\label{141958_24Feb21}
\end{equation}
Note that this is independent of $t$.
For the upper limit of $B_0$, yielding $\epsilon\sim 10^{-11}$, 
the induced current \eqref{141958_24Feb21} is larger than the maximum current estimated in the last subsection, 
$J_\mr{EM} \sim 2\pi v_s/e$, with $v_s \simeq 10^{9-12} \, \mr{GeV}$.
Therefore, the current is saturated to the maximum value by PMF.
In the following argument, we use this maximum current as those the strings carry in the early universe.

\begin{figure}[t]
\centering
\includegraphics[width=0.5\textwidth]{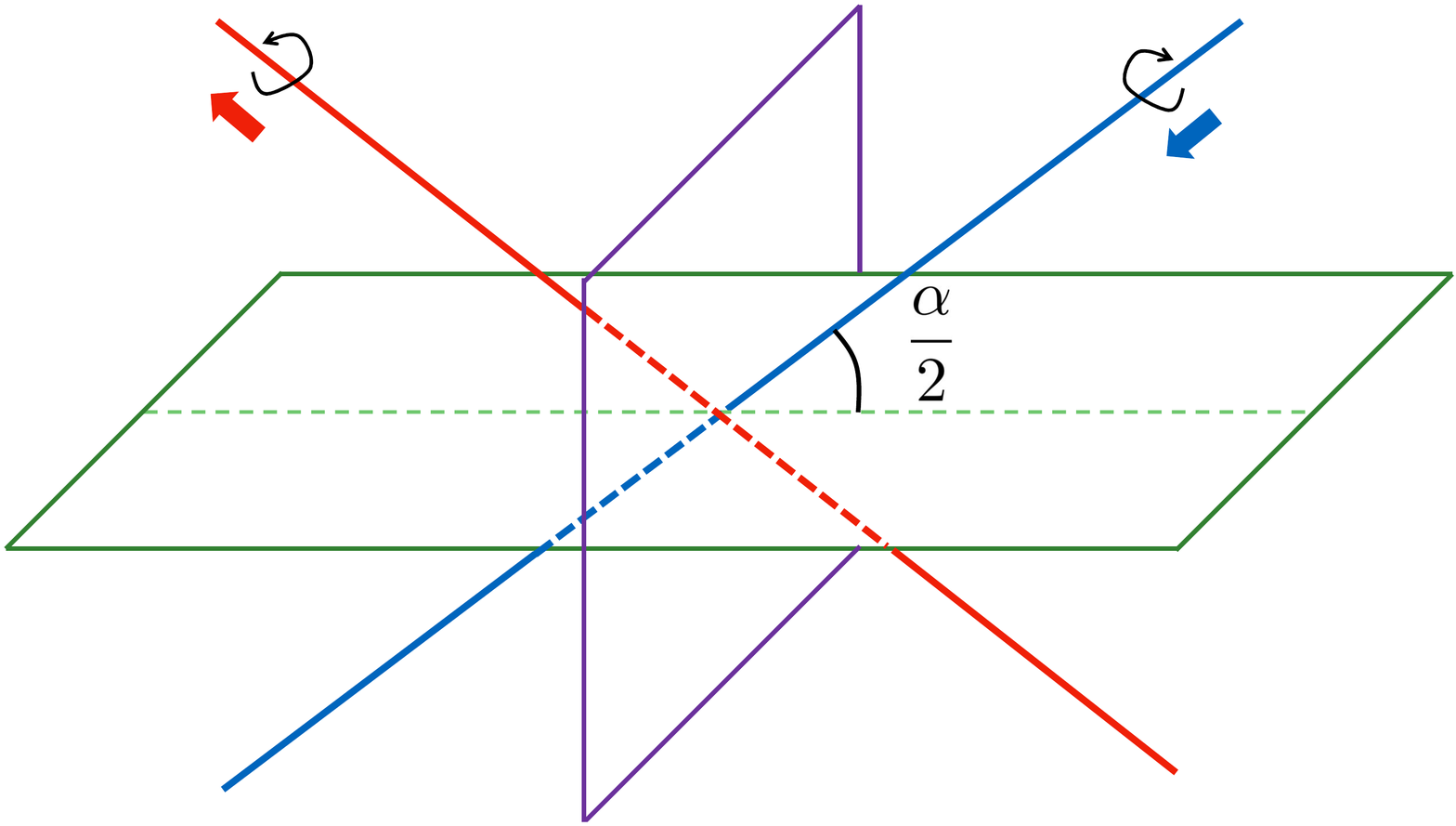}
\caption{
Two strings (red and blue lines) collide with the crossing angle $\alpha$.
The black arrows indicate the directions of the topological winding number of $U(1)_\mr{PQ}$.
The red and blue arrows indicate the directions of the flowing electric currents,
which are parallel for $\alpha=0$.
The green and purple planes are orthogonal and include the collision point.
}
\label{143148_24Feb21}
\end{figure}

Let us discuss the formation of Y-junctions.
They can be formed by collisions of two superconducting strings
when they feel an attractive force and are trapped in the potential.
We suppose that the two strings collide with the crossing angle $\alpha$
and that they are identical and parallel for $\alpha=0$.
The superconducting currents are assumed to flow in the same direction for $\alpha=0$ 
(the opposite case will be considered later).
The three-dimensional dynamics of the collision event is rather complicated and difficult to analyze.
However a qualitative picture can be understood by reducing it on two orthogonal planes including the colliding point \cite{Copeland:1986ng,Shellard:1987bv,Shellard:1988} (see Fig.~\ref{143148_24Feb21}).
On one of them (purple plane in Fig.~\ref{143148_24Feb21}), 
the situation is regarded as a collision (scattering) event of two point-like vortices in two dimensions
while, on the other plane (green plane in Fig.~\ref{143148_24Feb21}), 
as a vortex-antivortex collision resulting in the annihilation in two dimensions.
Note that this picture reproduces the reconnection process 
shown in the upper-right picture in Fig.~\ref{142848_24Feb21}
when the strings have no superconducting current;
two scattered vortices on the purple plane and nothing on the green one.

We concentrate on the two-dimensional analysis of the superconducting vortex-vortex collision
since the current does not change the annihilation.
Unlike the non-superconducting case leading to the $90^\circ$ scattering,
the long-range magnetic interaction plays a crucial role in our case.
On the reduced plane, the vortex-vortex pair feel a net interaction potential,
\begin{equation}
 V \sim \frac{v_s^2}{e^2}\cos \frac{\alpha}{2} \, \log r \, , \label{145301_24Feb21} 
\end{equation}
where $r$ is the distance between the two vortices
and we have ignored the axion-mediated repulsive interaction,
which should be subleading by a factor $e^2$ compared to Eq.~\eqref{145301_24Feb21}.
If the vortices do not have sufficient kinetic energy to escape to infinity, they become trapped by the potential.
Thus we obtain a condition to form a bound state,
\begin{equation}
 \frac{v_s^2}{e^2}\cos \frac{\alpha}{2} \, \log \frac{L}{\delta} \geq  (\gamma-1)\mu\label{155035_25Feb21} \, ,
\end{equation}
where $\gamma$ and $\mu$ are the Lorentz factor and the tension of the colliding vortices.
$L$ and $\delta$ are the IR and UV cutoff for the potential, 
which we take as the Hubble radius and the width of the strings, respectively.
As considering just after the electroweak phase transition,
the logarithmic factor gives $\log (10^{17} \times v_s/v_\mr{EW})\sim 50$.
In addition, we take a mildly relativistic velocity $\sim 0.6$ for the vortices, yielding $\gamma \sim 1.25$,
and thus Eq.~\eqref{155035_25Feb21} gives
\begin{equation}
\cos \frac{\alpha}{2} \geq  10^{-3} \, ,
\end{equation}
which means that they \textit{almost always} form the bound state except for $\alpha \simeq \pi$.

On the other hand, 
they cannot form such a bound state in the case that 
the currents flow in the opposite directions (anti-parallel) on the vortex-vortex plane
because they always feel a repulsive force.
Therefore, we obtain a rough estimation for the formation probability of the bound state, 
\begin{equation}
\begin{cases}
\text{Y-junction formed} & \text{(current in the same directions)} \\
\text{not formed} & \text{(current in the opposite directions)} 
\end{cases} 
\Rightarrow \text{probability} \simeq \frac{1}{2} \, .
\end{equation}
Note that this is based on the two-dimensional analysis focusing on the collision point.
This picture may break down for the case that the dynamics after colliding is dominated by other parts of the strings than the collision point.
Furthermore, the formed Y-junction could be peeled off, depending on the velocities and the crossing angle of the strings.
For a more detailed study including these effects, it is necessary to perform a three-dimensional simulation, which is beyond the scope of the present paper.

\begin{figure}[t]
\centering
\includegraphics[width=0.55\textwidth]{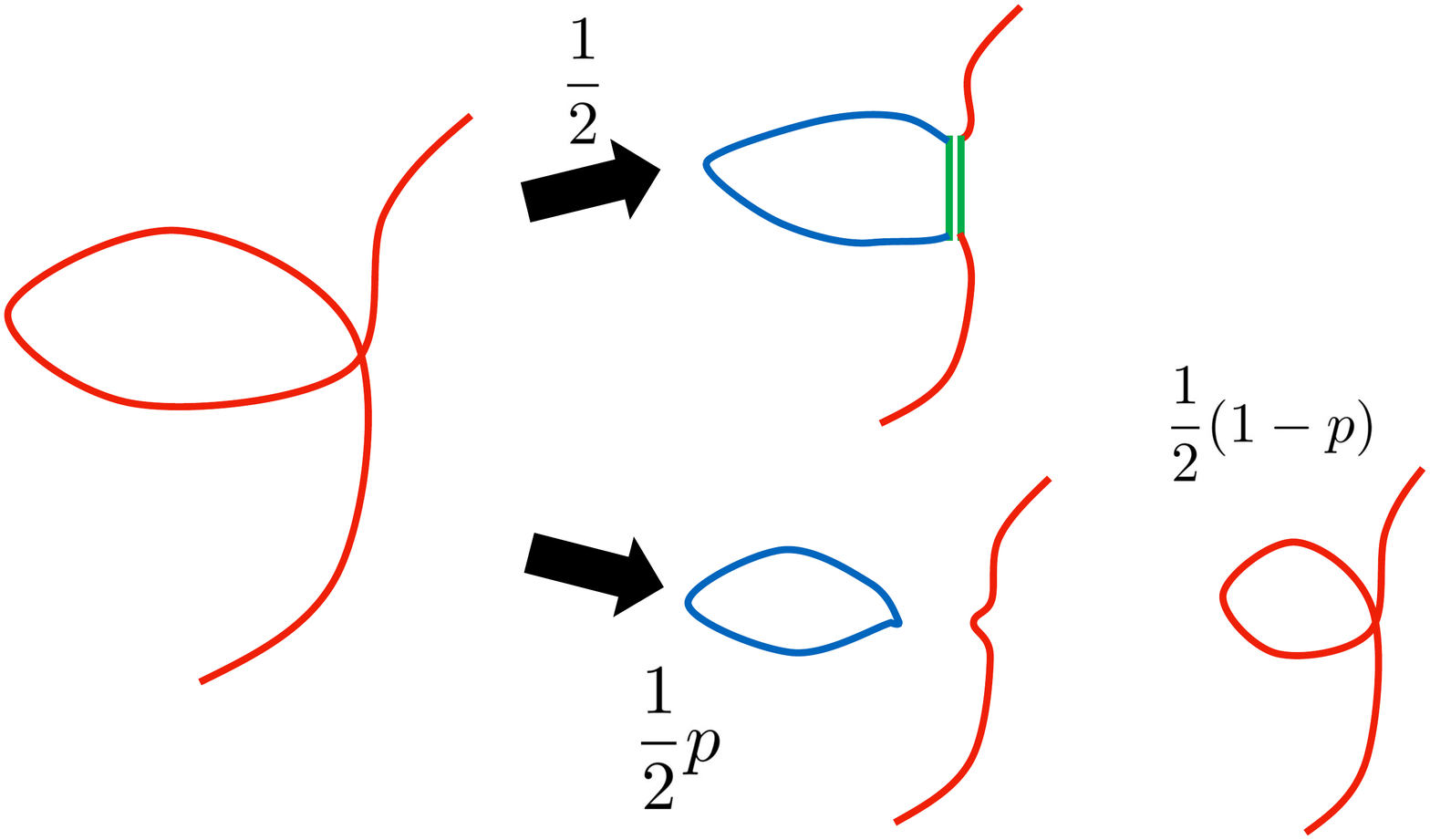}
\caption{
Self-intersection of a superconducting string.
The Y-junction (green doubled line) is formed with probability $\frac{1}{2}$.
Otherwise, it reconnects producing a small loop (blue line) or passes through.
The reconnection probability is $p\sim 1$.
}
\label{213808_25Feb21}
\end{figure}

Finally, we discuss a consequence of the formation of Y-junctions.
When a superconducting string intersects with itself, it forms the Y-junction with probability $1/2$,
produces a small loop with probability $p/2$,
or passes through with probability $(1-p)/2$ (see Fig.~\ref{213808_25Feb21}),
where $p$ is the reconnection probability without the current ($p \sim 1$).
If such a Y-junction-connected loop (upper one in Fig.~\ref{213808_25Feb21}) is produced,
it is not obvious whether the strings sufficiently loose their energy,
and the scaling behavior of the energy density of strings is not ensured.
Once the energy density of the universe is dominated by such superconducting strings, 
it causes large impacts on cosmology,
and hence some parameter region in the DFSZ axion model may be constrained.



\section{Conclusion}
\label{sec:conclusion}

We have studied the axion strings with the electroweak gauge flux,
and their superconductivity in the DFSZ model.
We constructed three types of the electroweak axion string solutions,
which have similar properties to
those of (non-Abelian) vortices in two Higgs doublet model.
We also showed that
in some parameter space,
the string with $W$-flux, we called the type-C string,
can be lighter than those with $Z$-flux.
The type-C string exhibits superconductivity
and a large electric current can flow along the string.
This large current may realize a net attractive force between the axion strings,
which could form the so-called Y-junctions in the early universe.
By considering the string collision reduced onto two-dimensional planes,
the probability of the formation of the Y-junctions is estimated to be $\sim 1/2$.

Once such Y-junctions are formed,
they can affect the evolution of the string network and
make non-trivial whether the network evolves to the scaling solution.
If no obeying the scaling behavior,
the string network could dominate the energy density of the the universe
and the model is severely constrained.
To conclude whether this is true or not, we need detailed numerical simulations on the time evolution of the network
taking into account both the axionic and magnetic interactions.
In addition, the Y-junction provides characteristic signals in astrophysical observations.
It is worthwhile to study the Y-junction dynamics of
the DFSZ axion strings\footnote{
Y-junctions could also be formed in the KSVZ model
when heavy extra fermions are $U(1)_\mathrm{EM}$ charged and the strings are superconducting.
}
within a phenomenologically allowed parameter space.

In this paper,
we focus on the $U(1)_{\mathrm{EM}}$ symmetry breaking
through the $W$-flux and the charged Higgs components.
On the other hand, the type-A and type-B strings also could be superconducting.
The $Z$-flux give a negative mass contribution to the $W$-boson as
\begin{align}
ig \cos \theta_W \ol{Z}_{\mu\nu}  
 W^{-,\mu} W^{+,\nu},
\end{align}
which leads the instability called the $W$-condensate \cite{Ambjorn:1992ca,Ambjorn:1989sz,Ambjorn:1989bd}.
This can be another possible mechanism realizing superconductivity.
If the magnitude of the background flux is sufficiently large,
the type-A and type-B strings also can be superconducting.

\section*{Note added}
After we completed the present work, 
an analysis on cosmological consequences of the superconducting axion string appeared in Ref.~\cite{Fukuda:2020kym},
in which the authors discussed the KSVZ model as a benchmark model.
As we stated in introduction, the DFSZ model has no heavy fermion,
and thus the supercurrent carried by fermionic zero modes is not significant compared to the string tension.
Our present argument that 
the string in the DFSZ model can carry a large amount of the supercurrent using the bosonic zero mode 
allows the studies in Ref.~\cite{Fukuda:2020kym} to be applied to the DFSZ model as well.
Thus our present work is complementary to theirs.


\section*{Acknowledgements}

\noindent
We would like to thank Minoru Eto and Muneto Nitta for valuable discussions on superconductivity of non-Abelian vortices.
We also thank Takahiro Ohata for discussion in the early stage of this work. 
This work is supported in part by JSPS Grant-in-Aid for Scientific Research
KAKENHI Grant No. JP18J22733 (Y.~H.),
No. JP18H01214, JP20K03949 (K.~Y.)
and
No. JP20J11901 (Y.~A.)
.

\appendix
\section{Derivation of the linearized EOMs}
\label{041710_25Sep20}
By substituting the string ansatz into the EOM~\eqref{160020_24Sep20} 
for $\mu=\alpha$, we have
\begin{equation}
 \left(\ol{D}_j  \ol{D}^j \delta W^\alpha \right)^a = 
\f{-i}{2}g ~\mathrm{Tr} \left[\ol{H} ^\dagger \sigma^a \left(-ig
    \delta W^\alpha \ol{H} + i\f{g'}{2} \delta Y^\alpha \ol{H}\sigma_3\right) 
- \text{h.c.} 
\right],
\end{equation}
where we have used the fact that
the background configurations $\ol{H}$, $\ol{W}_\mu$ and $\ol{Y}_\mu$
solve the EOMs. The equation is explicitly written by the 
functions $\eta$, $\chi$ and $\xi$ :
\begin{align}
  (\text{l.h.s.}) &=
\f{1}{g} \partial^\alpha \eta \left(\ol{D}_j  \ol{D}^j \chi \right)^a,  \\
 (\text{r.h.s}) & = \f{-g}{2} \partial^\alpha \eta ~\mathrm{Tr} \left[\ol{H} ^\dagger \left\{\sigma^a,\chi\right\} \ol{H}+
 \left(\ol{H}^\dagger \sigma_a \ol{H} \f{\sigma_3}{2} + \f{\sigma_3}{2} \ol{H}^\dagger \sigma_a \ol{H}\right) \xi \right]\\
& =\f{-g}{2} \partial^\alpha \eta ~ \left(\chi^a  \mathrm{Tr}|\ol{H}|^2
 + \xi \mathrm{Tr} \left[\ol{H}^\dagger \sigma^a \ol{H} \sigma_3  \right] \right),
\end{align}
which implies the linearized EOM \eqref{163119_24Sep20}. A similar
procedure for the EOM of $Y_\mu$ \eqref{165528_24Sep20} leads to
the linearized one \eqref{035453_25Sep20}.
On the other hand, for the EOM \eqref{160020_24Sep20} for $\mu=j$, we have
\begin{align}
&\left( D_\nu W ^{\nu j } \right)^a =
 - \f{i}{2}g ~\mathrm{Tr} \left[\ol{H} ^\dagger \sigma^a \ol{D} ^j \ol{H} - (\ol{D}^j \ol{H}) ^\dagger \sigma^a \ol{H}\right] .\label{163902_24Sep20}
\end{align}
The l.h.s.\ is divided into two pieces
\begin{align}
\left( D_\nu W ^{\nu j } \right)^a 
& = \left( \ol{D}_k W ^{k j } \right)^a + \left( D_\alpha W ^{\alpha j} \right)^a\label{163849_24Sep20}
\end{align}
with $k=r,\theta$.
The first term in the r.h.s.\ of Eq.~\eqref{163849_24Sep20} is found
to be equal to the r.h.s.\ of Eq.~\eqref{163902_24Sep20}, and they are
cancelled out from the EOM\@. The remaining second term is given at
the leading order of $\delta W$ as
\begin{align}
  D_\alpha W ^{\alpha j } &= 
-\partial_\alpha \ol{D}^j \delta W^\alpha + ig \left[\delta W_\alpha,\ol{D}^j \delta W^\alpha \right] 
\simeq  \f{-1}{g} \ol{D}^j \chi  ~\partial^\alpha \partial_\alpha \eta,
\end{align}
which implies the linearized EOM \eqref{164640_24Sep20}. The similar
derivation holds for the gauge field $Y_\mu$.


\bibliographystyle{jhep}
\bibliography{./references}

\providecommand{\href}[2]{#2}\begingroup\raggedright\begin{thebibliography}{10}

\bibitem{Peccei:1977ur}
R.~Peccei and H.~R. Quinn, \emph{{Constraints Imposed by CP Conservation in the
  Presence of Instantons}},
  \href{https://doi.org/10.1103/PhysRevD.16.1791}{\emph{Phys. Rev. D}
  {\bfseries 16} (1977) 1791}.

\bibitem{Peccei:1977hh}
R.~Peccei and H.~R. Quinn, \emph{{CP Conservation in the Presence of
  Instantons}}, \href{https://doi.org/10.1103/PhysRevLett.38.1440}{\emph{Phys.
  Rev. Lett.} {\bfseries 38} (1977) 1440}.

\bibitem{Weinberg:1977ma}
S.~Weinberg, \emph{{A New Light Boson?}},
  \href{https://doi.org/10.1103/PhysRevLett.40.223}{\emph{Phys. Rev. Lett.}
  {\bfseries 40} (1978) 223}.

\bibitem{Wilczek:1977pj}
F.~Wilczek, \emph{{Problem of Strong $P$ and $T$ Invariance in the Presence of
  Instantons}}, \href{https://doi.org/10.1103/PhysRevLett.40.279}{\emph{Phys.
  Rev. Lett.} {\bfseries 40} (1978) 279}.

\bibitem{Preskill:1982cy}
J.~Preskill, M.~B. Wise and F.~Wilczek, \emph{{Cosmology of the Invisible
  Axion}}, \href{https://doi.org/10.1016/0370-2693(83)90637-8}{\emph{Phys.
  Lett. B} {\bfseries 120} (1983) 127}.

\bibitem{Abbott:1982af}
L.~Abbott and P.~Sikivie, \emph{{A Cosmological Bound on the Invisible Axion}},
  \href{https://doi.org/10.1016/0370-2693(83)90638-X}{\emph{Phys. Lett. B}
  {\bfseries 120} (1983) 133}.

\bibitem{Dine:1982ah}
M.~Dine and W.~Fischler, \emph{{The Not So Harmless Axion}},
  \href{https://doi.org/10.1016/0370-2693(83)90639-1}{\emph{Phys. Lett. B}
  {\bfseries 120} (1983) 137}.

\bibitem{Sikivie:2006ni}
P.~Sikivie, \emph{{Axion Cosmology}},
  \href{https://doi.org/10.1007/978-3-540-73518-2_2}{\emph{Lect. Notes Phys.}
  {\bfseries 741} (2008) 19}
  [\href{https://arxiv.org/abs/astro-ph/0610440}{{\ttfamily
  astro-ph/0610440}}].

\bibitem{Marsh:2015xka}
D.~J.~E. Marsh, \emph{{Axion Cosmology}},
  \href{https://doi.org/10.1016/j.physrep.2016.06.005}{\emph{Phys. Rept.}
  {\bfseries 643} (2016) 1} [\href{https://arxiv.org/abs/1510.07633}{{\ttfamily
  1510.07633}}].

\bibitem{Ringwald:2012hr}
A.~Ringwald, \emph{{Exploring the Role of Axions and Other WISPs in the Dark
  Universe}}, \href{https://doi.org/10.1016/j.dark.2012.10.008}{\emph{Phys.
  Dark Univ.} {\bfseries 1} (2012) 116}
  [\href{https://arxiv.org/abs/1210.5081}{{\ttfamily 1210.5081}}].

\bibitem{Wantz:2009it}
O.~Wantz and E.~Shellard, \emph{{Axion Cosmology Revisited}},
  \href{https://doi.org/10.1103/PhysRevD.82.123508}{\emph{Phys. Rev. D}
  {\bfseries 82} (2010) 123508}
  [\href{https://arxiv.org/abs/0910.1066}{{\ttfamily 0910.1066}}].

\bibitem{Kim:2008hd}
J.~E. Kim and G.~Carosi, \emph{{Axions and the Strong CP Problem}},
  \href{https://doi.org/10.1103/RevModPhys.82.557}{\emph{Rev. Mod. Phys.}
  {\bfseries 82} (2010) 557} [\href{https://arxiv.org/abs/0807.3125}{{\ttfamily
  0807.3125}}].

\bibitem{Zhitnitsky:1980tq}
A.~Zhitnitsky, \emph{{On Possible Suppression of the Axion Hadron Interactions.
  (In Russian)}}, {\emph{Sov. J. Nucl. Phys.} {\bfseries 31} (1980) 260}.

\bibitem{Dine:1981rt}
M.~Dine, W.~Fischler and M.~Srednicki, \emph{{A Simple Solution to the Strong
  CP Problem with a Harmless Axion}},
  \href{https://doi.org/10.1016/0370-2693(81)90590-6}{\emph{Phys. Lett. B}
  {\bfseries 104} (1981) 199}.

\bibitem{Kim:1979if}
J.~E. Kim, \emph{{Weak Interaction Singlet and Strong CP Invariance}},
  \href{https://doi.org/10.1103/PhysRevLett.43.103}{\emph{Phys. Rev. Lett.}
  {\bfseries 43} (1979) 103}.

\bibitem{Shifman:1979if}
M.~A. Shifman, A.~Vainshtein and V.~I. Zakharov, \emph{{Can Confinement Ensure
  Natural CP Invariance of Strong Interactions?}},
  \href{https://doi.org/10.1016/0550-3213(80)90209-6}{\emph{Nucl. Phys. B}
  {\bfseries 166} (1980) 493}.

\bibitem{Akrami:2018odb}
{\scshape Planck} collaboration, \emph{{Planck 2018 results. X. Constraints on
  inflation}}, \href{https://doi.org/10.1051/0004-6361/201833887}{\emph{Astron.
  Astrophys.} {\bfseries 641} (2020) A10}
  [\href{https://arxiv.org/abs/1807.06211}{{\ttfamily 1807.06211}}].

\bibitem{Gelmini:1988sf}
G.~B. Gelmini, M.~Gleiser and E.~W. Kolb, \emph{{Cosmology of Biased Discrete
  Symmetry Breaking}},
  \href{https://doi.org/10.1103/PhysRevD.39.1558}{\emph{Phys. Rev. D}
  {\bfseries 39} (1989) 1558}.

\bibitem{Larsson:1996sp}
S.~E. Larsson, S.~Sarkar and P.~L. White, \emph{{Evading the cosmological
  domain wall problem}},
  \href{https://doi.org/10.1103/PhysRevD.55.5129}{\emph{Phys. Rev. D}
  {\bfseries 55} (1997) 5129}
  [\href{https://arxiv.org/abs/hep-ph/9608319}{{\ttfamily hep-ph/9608319}}].

\bibitem{Sikivie:1982qv}
P.~Sikivie, \emph{{Of Axions, Domain Walls and the Early Universe}},
  \href{https://doi.org/10.1103/PhysRevLett.48.1156}{\emph{Phys. Rev. Lett.}
  {\bfseries 48} (1982) 1156}.

\bibitem{Chang:1998tb}
S.~Chang, C.~Hagmann and P.~Sikivie, \emph{{Studies of the motion and decay of
  axion walls bounded by strings}},
  \href{https://doi.org/10.1103/PhysRevD.59.023505}{\emph{Phys. Rev. D}
  {\bfseries 59} (1999) 023505}
  [\href{https://arxiv.org/abs/hep-ph/9807374}{{\ttfamily hep-ph/9807374}}].

\bibitem{Vilenkin:1981zs}
A.~Vilenkin, \emph{{Gravitational Field of Vacuum Domain Walls and Strings}},
  \href{https://doi.org/10.1103/PhysRevD.23.852}{\emph{Phys. Rev. D} {\bfseries
  23} (1981) 852}.

\bibitem{Peccei:1986pn}
R.~Peccei, T.~T. Wu and T.~Yanagida, \emph{{A viable axion model}},
  \href{https://doi.org/10.1016/0370-2693(86)90284-4}{\emph{Phys. Lett. B}
  {\bfseries 172} (1986) 435}.

\bibitem{Krauss:1986wx}
L.~M. Krauss and F.~Wilczek, \emph{{A shortlived axion variant}},
  \href{https://doi.org/10.1016/0370-2693(86)90244-3}{\emph{Phys. Lett. B}
  {\bfseries 173} (1986) 189}.

\bibitem{Lazarides:1982tw}
G.~Lazarides and Q.~Shafi, \emph{{Axion Models with No Domain Wall Problem}},
  \href{https://doi.org/10.1016/0370-2693(82)90506-8}{\emph{Phys. Lett. B}
  {\bfseries 115} (1982) 21}.

\bibitem{Chatterjee:2019rch}
C.~Chatterjee, T.~Higaki and M.~Nitta, \emph{{Note on a solution to domain wall
  problem with the Lazarides-Shafi mechanism in axion dark matter models}},
  \href{https://doi.org/10.1103/PhysRevD.101.075026}{\emph{Phys. Rev. D}
  {\bfseries 101} (2020) 075026}
  [\href{https://arxiv.org/abs/1903.11753}{{\ttfamily 1903.11753}}].

\bibitem{Kawasaki:2015lpf}
M.~Kawasaki, F.~Takahashi and M.~Yamada, \emph{{Suppressing the QCD Axion
  Abundance by Hidden Monopoles}},
  \href{https://doi.org/10.1016/j.physletb.2015.12.075}{\emph{Phys. Lett. B}
  {\bfseries 753} (2016) 677}
  [\href{https://arxiv.org/abs/1511.05030}{{\ttfamily 1511.05030}}].

\bibitem{Sato:2018nqy}
R.~Sato, F.~Takahashi and M.~Yamada, \emph{{Unified Origin of Axion and
  Monopole Dark Matter, and Solution to the Domain-wall Problem}},
  \href{https://doi.org/10.1103/PhysRevD.98.043535}{\emph{Phys. Rev. D}
  {\bfseries 98} (2018) 043535}
  [\href{https://arxiv.org/abs/1805.10533}{{\ttfamily 1805.10533}}].

\bibitem{Davis:1986xc}
R.~L. Davis, \emph{{Cosmic Axions from Cosmic Strings}},
  \href{https://doi.org/10.1016/0370-2693(86)90300-X}{\emph{Phys. Lett. B}
  {\bfseries 180} (1986) 225}.

\bibitem{Kibble:1980mv}
T.~W.~B. Kibble, \emph{{Some Implications of a Cosmological Phase Transition}},
  \href{https://doi.org/10.1016/0370-1573(80)90091-5}{\emph{Phys. Rept.}
  {\bfseries 67} (1980) 183}.

\bibitem{Zurek:1985qw}
W.~H. Zurek, \emph{{Cosmological Experiments in Superfluid Helium?}},
  \href{https://doi.org/10.1038/317505a0}{\emph{Nature} {\bfseries 317} (1985)
  505}.

\bibitem{Murayama:2009nj}
H.~Murayama and J.~Shu, \emph{{Topological Dark Matter}},
  \href{https://doi.org/10.1016/j.physletb.2010.02.037}{\emph{Phys. Lett. B}
  {\bfseries 686} (2010) 162}
  [\href{https://arxiv.org/abs/0905.1720}{{\ttfamily 0905.1720}}].

\bibitem{Vilenkin:2000jqa}
A.~Vilenkin and E.~S. Shellard, \emph{{Cosmic Strings and Other Topological
  Defects}}. Cambridge University Press, 7, 2000.

\bibitem{Witten:1984eb}
E.~Witten, \emph{{Superconducting Strings}},
  \href{https://doi.org/10.1016/0550-3213(85)90022-7}{\emph{Nucl. Phys. B}
  {\bfseries 249} (1985) 557}.

\bibitem{Lazarides:1984zq}
G.~Lazarides and Q.~Shafi, \emph{{Superconducting Strings in Axion Models}},
  \href{https://doi.org/10.1016/0370-2693(85)91398-X}{\emph{Phys. Lett. B}
  {\bfseries 151} (1985) 123}.

\bibitem{Iwazaki:1997bk}
A.~Iwazaki, \emph{{Spontaneous magnetization of axion domain wall and
  primordial magnetic field}},
  \href{https://doi.org/10.1103/PhysRevLett.79.2927}{\emph{Phys. Rev. Lett.}
  {\bfseries 79} (1997) 2927}
  [\href{https://arxiv.org/abs/hep-ph/9705456}{{\ttfamily hep-ph/9705456}}].

\bibitem{Ganoulis:1989hz}
N.~Ganoulis and G.~Lazarides, \emph{{Fermionic Zero Modes for Cosmic Strings}},
  \href{https://doi.org/10.1016/0550-3213(89)90040-0}{\emph{Nucl. Phys. B}
  {\bfseries 316} (1989) 443}.

\bibitem{Lazarides:1987rq}
G.~Lazarides, C.~Panagiotakopoulos and Q.~Shafi, \emph{{COSMIC SUPERCONDUCTING
  STRINGS AND COLLIDERS}},
  \href{https://doi.org/10.1016/0550-3213(88)90037-5}{\emph{Nucl. Phys. B}
  {\bfseries 296} (1988) 657}.

\bibitem{Jackiw:1981ee}
R.~Jackiw and P.~Rossi, \emph{{Zero Modes of the Vortex - Fermion System}},
  \href{https://doi.org/10.1016/0550-3213(81)90044-4}{\emph{Nucl. Phys. B}
  {\bfseries 190} (1981) 681}.

\bibitem{Callan:1984sa}
J.~Callan, Curtis~G. and J.~A. Harvey, \emph{{Anomalies and Fermion Zero Modes
  on Strings and Domain Walls}},
  \href{https://doi.org/10.1016/0550-3213(85)90489-4}{\emph{Nucl. Phys. B}
  {\bfseries 250} (1985) 427}.

\bibitem{Abrikosov:1956sx}
A.~A. Abrikosov, \emph{{On the Magnetic properties of superconductors of the
  second group}}, {\emph{Sov. Phys. JETP} {\bfseries 5} (1957) 1174}.

\bibitem{Nielsen:1973cs}
H.~B. Nielsen and P.~Olesen, \emph{{Vortex Line Models for Dual Strings}},
  \href{https://doi.org/10.1016/0550-3213(73)90350-7}{\emph{Nucl. Phys.}
  {\bfseries B61} (1973) 45}.

\bibitem{Nambu:1977ag}
Y.~Nambu, \emph{{String-Like Configurations in the Weinberg-Salam Theory}},
  \href{https://doi.org/10.1016/0550-3213(77)90252-8}{\emph{Nucl. Phys.}
  {\bfseries B130} (1977) 505}.

\bibitem{Vachaspati:1992jk}
T.~Vachaspati, \emph{{Electroweak strings}},
  \href{https://doi.org/10.1016/0550-3213(93)90189-V}{\emph{Nucl. Phys.}
  {\bfseries B397} (1993) 648}.

\bibitem{Vachaspati:1992fi}
T.~Vachaspati, \emph{{Vortex solutions in the Weinberg-Salam model}},
  \href{https://doi.org/10.1103/PhysRevLett.68.1977,
  10.1103/PhysRevLett.69.216.2}{\emph{Phys. Rev. Lett.} {\bfseries 68} (1992)
  1977}.

\bibitem{James:1992zp}
M.~James, L.~Perivolaropoulos and T.~Vachaspati, \emph{{Stability of
  electroweak strings}},
  \href{https://doi.org/10.1103/PhysRevD.46.R5232}{\emph{Phys. Rev.} {\bfseries
  D46} (1992) R5232}.

\bibitem{James:1992wb}
M.~James, L.~Perivolaropoulos and T.~Vachaspati, \emph{{Detailed stability
  analysis of electroweak strings}},
  \href{https://doi.org/10.1016/0550-3213(93)90046-R}{\emph{Nucl. Phys.}
  {\bfseries B395} (1993) 534}
  [\href{https://arxiv.org/abs/hep-ph/9212301}{{\ttfamily hep-ph/9212301}}].

\bibitem{Vachaspati:1994ng}
T.~Vachaspati and G.~B. Field, \emph{{Electroweak string configurations with
  baryon number}},
  \href{https://doi.org/10.1103/PhysRevLett.73.373}{\emph{Phys. Rev. Lett.}
  {\bfseries 73} (1994) 373}
  [\href{https://arxiv.org/abs/hep-ph/9401220}{{\ttfamily hep-ph/9401220}}].

\bibitem{Barriola:1993fy}
M.~Barriola, T.~Vachaspati and M.~Bucher, \emph{{Embedded defects}},
  \href{https://doi.org/10.1103/PhysRevD.50.2819}{\emph{Phys. Rev.} {\bfseries
  D50} (1994) 2819} [\href{https://arxiv.org/abs/hep-th/9306120}{{\ttfamily
  hep-th/9306120}}].

\bibitem{Barriola:1994ez}
M.~Barriola, \emph{{Electroweak strings that produce baryons}},
  \href{https://doi.org/10.1103/PhysRevD.51.R300}{\emph{Phys. Rev.} {\bfseries
  D51} (1995) 300} [\href{https://arxiv.org/abs/hep-ph/9403323}{{\ttfamily
  hep-ph/9403323}}].

\bibitem{Eto:2012kb}
M.~Eto, K.~Konishi, M.~Nitta and Y.~Ookouchi, \emph{{Brane Realization of Nambu
  Monopoles and Electroweak Strings}},
  \href{https://doi.org/10.1103/PhysRevD.87.045006}{\emph{Phys. Rev.}
  {\bfseries D87} (2013) 045006}
  [\href{https://arxiv.org/abs/1211.2971}{{\ttfamily 1211.2971}}].

\bibitem{Achucarro:1999it}
A.~Achucarro and T.~Vachaspati, \emph{{Semilocal and electroweak strings}},
  \href{https://doi.org/10.1016/S0370-1573(99)00103-9}{\emph{Phys. Rept.}
  {\bfseries 327} (2000) 347}
  [\href{https://arxiv.org/abs/hep-ph/9904229}{{\ttfamily hep-ph/9904229}}].

\bibitem{Perivolaropoulos:1993gg}
L.~Perivolaropoulos, \emph{{Existence of double vortex solutions}},
  \href{https://doi.org/10.1016/0370-2693(93)91039-P}{\emph{Phys. Lett.}
  {\bfseries B316} (1993) 528}
  [\href{https://arxiv.org/abs/hep-ph/9309261}{{\ttfamily hep-ph/9309261}}].

\bibitem{La:1993je}
H.~La, \emph{{Vortex solutions in two Higgs systems and tan Beta}},
  \href{https://arxiv.org/abs/hep-ph/9302220}{{\ttfamily hep-ph/9302220}}.

\bibitem{Dvali:1993sg}
G.~R. Dvali and G.~Senjanovic, \emph{{Topologically stable electroweak flux
  tubes}}, \href{https://doi.org/10.1103/PhysRevLett.71.2376}{\emph{Phys. Rev.
  Lett.} {\bfseries 71} (1993) 2376}
  [\href{https://arxiv.org/abs/hep-ph/9305278}{{\ttfamily hep-ph/9305278}}].

\bibitem{Dvali:1994qf}
G.~R. Dvali and G.~Senjanovic, \emph{{Topologically stable Z strings in the
  supersymmetric Standard Model}},
  \href{https://doi.org/10.1016/0370-2693(94)90943-1}{\emph{Phys. Lett.}
  {\bfseries B331} (1994) 63}
  [\href{https://arxiv.org/abs/hep-ph/9403277}{{\ttfamily hep-ph/9403277}}].

\bibitem{Bimonte:1994qh}
G.~Bimonte and G.~Lozano, \emph{{Vortex solutions in two Higgs doublet
  systems}}, \href{https://doi.org/10.1016/0370-2693(94)91321-8}{\emph{Phys.
  Lett.} {\bfseries B326} (1994) 270}
  [\href{https://arxiv.org/abs/hep-ph/9401313}{{\ttfamily hep-ph/9401313}}].

\bibitem{Bachas:1998bf}
C.~Bachas, B.~Rai and T.~N. Tomaras, \emph{{New string excitations in the two
  Higgs standard model}},
  \href{https://doi.org/10.1103/PhysRevLett.82.2443}{\emph{Phys. Rev. Lett.}
  {\bfseries 82} (1999) 2443}
  [\href{https://arxiv.org/abs/hep-ph/9801263}{{\ttfamily hep-ph/9801263}}].

\bibitem{Ivanov:2007de}
I.~P. Ivanov, \emph{{Minkowski space structure of the Higgs potential in 2HDM.
  II. Minima, symmetries, and topology}},
  \href{https://doi.org/10.1103/PhysRevD.77.015017}{\emph{Phys. Rev.}
  {\bfseries D77} (2008) 015017}
  [\href{https://arxiv.org/abs/0710.3490}{{\ttfamily 0710.3490}}].

\bibitem{Battye:2011jj}
R.~A. Battye, G.~D. Brawn and A.~Pilaftsis, \emph{{Vacuum Topology of the Two
  Higgs Doublet Model}},
  \href{https://doi.org/10.1007/JHEP08(2011)020}{\emph{JHEP} {\bfseries 08}
  (2011) 020} [\href{https://arxiv.org/abs/1106.3482}{{\ttfamily 1106.3482}}].

\bibitem{Eto:2019hhf}
M.~Eto, Y.~Hamada, M.~Kurachi and M.~Nitta, \emph{{Topological Nambu monopole
  in two Higgs doublet models}},
  \href{https://doi.org/10.1016/j.physletb.2020.135220}{\emph{Phys. Lett.}
  {\bfseries B802} (2020) 135220}
  [\href{https://arxiv.org/abs/1904.09269}{{\ttfamily 1904.09269}}].

\bibitem{Eto:2020hjb}
M.~Eto, Y.~Hamada, M.~Kurachi and M.~Nitta, \emph{{Dynamics of Nambu monopole
  in two Higgs doublet models. Cosmological Monopole Collider}},
  \href{https://doi.org/10.1007/JHEP07(2020)004}{\emph{JHEP} {\bfseries 07}
  (2020) 004} [\href{https://arxiv.org/abs/2003.08772}{{\ttfamily
  2003.08772}}].

\bibitem{Eto:2018hhg}
M.~Eto, M.~Kurachi and M.~Nitta, \emph{{Constraints on two Higgs doublet models
  from domain walls}},
  \href{https://doi.org/10.1016/j.physletb.2018.09.002}{\emph{Phys. Lett.}
  {\bfseries B785} (2018) 447}
  [\href{https://arxiv.org/abs/1803.04662}{{\ttfamily 1803.04662}}].

\bibitem{Eto:2018tnk}
M.~Eto, M.~Kurachi and M.~Nitta, \emph{{Non-Abelian strings and domain walls in
  two Higgs doublet models}},
  \href{https://doi.org/10.1007/JHEP08(2018)195}{\emph{JHEP} {\bfseries 08}
  (2018) 195} [\href{https://arxiv.org/abs/1805.07015}{{\ttfamily
  1805.07015}}].

\bibitem{Alford:1990mk}
M.~G. Alford, K.~Benson, S.~R. Coleman, J.~March-Russell and F.~Wilczek,
  \emph{{The Interactions and Excitations of Nonabelian Vortices}},
  \href{https://doi.org/10.1103/PhysRevLett.64.1632}{\emph{Phys. Rev. Lett.}
  {\bfseries 64} (1990) 1632}.

\bibitem{Alford:1990ur}
M.~G. Alford, K.~Benson, S.~R. Coleman, J.~March-Russell and F.~Wilczek,
  \emph{{Zero modes of nonabelian vortices}},
  \href{https://doi.org/10.1016/0550-3213(91)90331-Q}{\emph{Nucl. Phys. B}
  {\bfseries 349} (1991) 414}.

\bibitem{Grzadkowski:2010dj}
B.~Grzadkowski, M.~Maniatis and J.~Wudka, \emph{{The bilinear formalism and the
  custodial symmetry in the two-Higgs-doublet model}},
  \href{https://doi.org/10.1007/JHEP11(2011)030}{\emph{JHEP} {\bfseries 11}
  (2011) 030} [\href{https://arxiv.org/abs/1011.5228}{{\ttfamily 1011.5228}}].

\bibitem{Pomarol:1993mu}
A.~Pomarol and R.~Vega, \emph{{Constraints on CP violation in the Higgs sector
  from the rho parameter}},
  \href{https://doi.org/10.1016/0550-3213(94)90611-4}{\emph{Nucl. Phys.}
  {\bfseries B413} (1994) 3}
  [\href{https://arxiv.org/abs/hep-ph/9305272}{{\ttfamily hep-ph/9305272}}].

\bibitem{Eto:2020opf}
M.~Eto, Y.~Hamada and M.~Nitta, \emph{{Topological structure of Nambu monopole
  in Two Higgs doublet models -- Fiber bundle, Dirac's quantization and dyon
  --}},  \href{https://arxiv.org/abs/2007.15587}{{\ttfamily 2007.15587}}.

\bibitem{Bettencourt:1996qe}
L.~M. Bettencourt, P.~Laguna and R.~A. Matzner, \emph{{Nonintercommuting cosmic
  strings}}, \href{https://doi.org/10.1103/PhysRevLett.78.2066}{\emph{Phys.
  Rev. Lett.} {\bfseries 78} (1997) 2066}
  [\href{https://arxiv.org/abs/hep-ph/9612350}{{\ttfamily hep-ph/9612350}}].

\bibitem{Bettencourt:1994kc}
L.~Bettencourt and T.~Kibble, \emph{{Nonintercommuting configurations in the
  collisions of type I U(1) cosmic strings}},
  \href{https://doi.org/10.1016/0370-2693(94)91257-2}{\emph{Phys. Lett. B}
  {\bfseries 332} (1994) 297}
  [\href{https://arxiv.org/abs/hep-ph/9405221}{{\ttfamily hep-ph/9405221}}].

\bibitem{Copeland:2006eh}
E.~Copeland, T.~Kibble and D.~A. Steer, \emph{{Collisions of strings with Y
  junctions}}, \href{https://doi.org/10.1103/PhysRevLett.97.021602}{\emph{Phys.
  Rev. Lett.} {\bfseries 97} (2006) 021602}
  [\href{https://arxiv.org/abs/hep-th/0601153}{{\ttfamily hep-th/0601153}}].

\bibitem{Copeland:2006if}
E.~Copeland, T.~Kibble and D.~A. Steer, \emph{{Constraints on string networks
  with junctions}},
  \href{https://doi.org/10.1103/PhysRevD.75.065024}{\emph{Phys. Rev. D}
  {\bfseries 75} (2007) 065024}
  [\href{https://arxiv.org/abs/hep-th/0611243}{{\ttfamily hep-th/0611243}}].

\bibitem{Salmi:2007ah}
P.~Salmi, A.~Achucarro, E.~Copeland, T.~Kibble, R.~de~Putter and D.~A. Steer,
  \emph{{Kinematic constraints on formation of bound states of cosmic strings:
  Field theoretical approach}},
  \href{https://doi.org/10.1103/PhysRevD.77.041701}{\emph{Phys. Rev. D}
  {\bfseries 77} (2008) 041701}
  [\href{https://arxiv.org/abs/0712.1204}{{\ttfamily 0712.1204}}].

\bibitem{Bevis:2008hg}
N.~Bevis and P.~M. Saffin, \emph{{Cosmic string Y-junctions: A Comparison
  between field theoretic and Nambu-Goto dynamics}},
  \href{https://doi.org/10.1103/PhysRevD.78.023503}{\emph{Phys. Rev. D}
  {\bfseries 78} (2008) 023503}
  [\href{https://arxiv.org/abs/0804.0200}{{\ttfamily 0804.0200}}].

\bibitem{Bevis:2009az}
N.~Bevis, E.~J. Copeland, P.-Y. Martin, G.~Niz, A.~Pourtsidou, P.~M. Saffin
  et~al., \emph{{Evolution and stability of cosmic string loops with
  Y-junctions}}, \href{https://doi.org/10.1103/PhysRevD.80.125030}{\emph{Phys.
  Rev. D} {\bfseries 80} (2009) 125030}
  [\href{https://arxiv.org/abs/0904.2127}{{\ttfamily 0904.2127}}].

\bibitem{Hiramatsu:2013yxa}
T.~Hiramatsu, M.~Eto, K.~Kamada, T.~Kobayashi and Y.~Ookouchi,
  \emph{{Instability of colliding metastable strings}},
  \href{https://doi.org/10.1007/JHEP01(2014)165}{\emph{JHEP} {\bfseries 01}
  (2014) 165} [\href{https://arxiv.org/abs/1304.0623}{{\ttfamily 1304.0623}}].

\bibitem{Hiramatsu:2013tga}
T.~Hiramatsu, Y.~Sendouda, K.~Takahashi, D.~Yamauchi and C.-M. Yoo,
  \emph{{Type-I cosmic string network}},
  \href{https://doi.org/10.1103/PhysRevD.88.085021}{\emph{Phys. Rev. D}
  {\bfseries 88} (2013) 085021}
  [\href{https://arxiv.org/abs/1307.0308}{{\ttfamily 1307.0308}}].

\bibitem{Parker:1955zz}
E.~N. Parker, \emph{{Hydromagnetic Dynamo Models}},
  \href{https://doi.org/10.1086/146087}{\emph{Astrophys. J.} {\bfseries 122}
  (1955) 293}.

\bibitem{Durrer:2013pga}
R.~Durrer and A.~Neronov, \emph{{Cosmological Magnetic Fields: Their
  Generation, Evolution and Observation}},
  \href{https://doi.org/10.1007/s00159-013-0062-7}{\emph{Astron. Astrophys.
  Rev.} {\bfseries 21} (2013) 62}
  [\href{https://arxiv.org/abs/1303.7121}{{\ttfamily 1303.7121}}].

\bibitem{Neronov_2010}
A.~Neronov and I.~Vovk, \emph{Evidence for strong extragalactic magnetic fields
  from fermi observations of tev blazars},
  \href{https://doi.org/10.1126/science.1184192}{\emph{Science} {\bfseries 328}
  (2010) 73^^e2^^80^^9375}.

\bibitem{Jedamzik_2019}
K.~Jedamzik and A.~Saveliev, \emph{Stringent limit on primordial magnetic
  fields from the cosmic microwave background radiation},
  \href{https://doi.org/10.1103/physrevlett.123.021301}{\emph{Physical Review
  Letters} {\bfseries 123} (2019) }.

\bibitem{Copeland:1986ng}
E.~J. Copeland and N.~Turok, \emph{{Cosmic String Interactions}}, .

\bibitem{Shellard:1987bv}
E.~P.~S. Shellard, \emph{{Cosmic String Interactions}},
  \href{https://doi.org/10.1016/0550-3213(87)90290-2}{\emph{Nucl. Phys. B}
  {\bfseries 283} (1987) 624}.

\bibitem{Shellard:1988}
E.~P.~S. Shellard, \emph{{Understanding intercommuting}}, {\emph{Proceedings of
  Yale Workshop: Cosmic Strings: The Current Status} (1988) }.

\bibitem{Ambjorn:1992ca}
J.~Ambjorn and P.~Olesen, \emph{{Electroweak magnetism, W condensation and
  antiscreening}},  in \emph{{4th Hellenic School on Elementary Particle
  Physics}}, pp.~396--406, 9, 1992,
  \href{https://arxiv.org/abs/hep-ph/9304220}{{\ttfamily hep-ph/9304220}}.

\bibitem{Ambjorn:1989sz}
J.~Ambjorn and P.~Olesen, \emph{{Electroweak Magnetism: Theory and
  Application}}, \href{https://doi.org/10.1142/S0217751X90001914}{\emph{Int. J.
  Mod. Phys. A} {\bfseries 5} (1990) 4525}.

\bibitem{Ambjorn:1989bd}
J.~Ambjorn and P.~Olesen, \emph{{A Condensate Solution of the Electroweak
  Theory Which Interpolates Between the Broken and the Symmetric Phase}},
  \href{https://doi.org/10.1016/0550-3213(90)90307-Y}{\emph{Nucl. Phys. B}
  {\bfseries 330} (1990) 193}.

\bibitem{Fukuda:2020kym}
H.~Fukuda, A.~V. Manohar, H.~Murayama and O.~Telem, \emph{{Axion strings are
  superconducting}},  \href{https://arxiv.org/abs/2010.02763}{{\ttfamily
  2010.02763}}.

\end{thebibliography}\endgroup

\end{document}